%% file: 2500d_clusters.tex
\documentclass[preprint2]{emulateapj} 
\usepackage{amsmath}
\usepackage{natbib}
\usepackage{graphicx}
\input{epsf}
\usepackage{epstopdf}
\usepackage{multirow}
\usepackage{epsfig}
\usepackage{color}
\usepackage{url}
\usepackage{verbatim}
\usepackage{subfigure}

\DeclareGraphicsExtensions{.jpg,.pdf,.png,.eps,.ps}

\newcommand{\msun}{\ensuremath{M_\odot}}
\newcommand{\Tcmb}{\mbox{$T_{\mbox{\tiny CMB}}$}}
\newcommand{\mass}{\mbox{$M_{\mbox{\scriptsize 500c}}$}}

\newcommand{\degs}{deg$^2$}
\newcommand{\muksq}{\ensuremath{\mu {\rm K}^2}}
\newcommand{\muk}{\ensuremath{\mu {\rm K}}}

\newcommand{\chandra}{{\sl Chandra}}
\newcommand{\spitzer}{{\sl Spitzer}}

\newcommand{\planck}{{\sl Planck}}
\newcommand{\rosat}{{\sl ROSAT}}

\newcommand{\Kcmb}{\mbox{$\mathrm{K}_{\mbox{\tiny CMB}}$}}
\newcommand{\um}{$\mu$m}

\newcommand{\nfalsefourf}{172}  
\newcommand{\nfalsefive}{18.5} 
\newcommand{\ncand}{677}  
\newcommand{\ncandfive}{409}  
\newcommand{\nconfirm}{516}  
\newcommand{\pctconfirm}{76}  
\newcommand{\nfourpseven}{530} 
\newcommand{\nlow}{147}  
\newcommand{\nconfirmfive}{387} 
\newcommand{\pctconfirmfive}{95} 
\newcommand{\purityfive}{95\%}  
\newcommand{\purityfourfive}{76\%} 
\newcommand{\nspec}{141}              
\newcommand{\nspecnew}{34}              
\newcommand{\ncandpsmask}{19}  
\newcommand{\medianz}{0.55}
\newcommand{\medianm}{\ensuremath{\sim3.5\times10^{14}\,M_\odot\,h_{70}^{-1}}}

\newcommand{\areamasked}{163}
\newcommand{\areasearched}{2365}

\newcommand{\scatterspitzer}{0.035}  
\newcommand{\sptcluster}{415}  
\newcommand{\newcluster}{251}   
\newcommand{\newclusterpercent}{49\%} 

\newcommand{\nSN}{\ensuremath{\xi}}

\newcommand{\fsz}{\ensuremath{f_\mathrm{\mbox{\tiny{SZ}}}}}
\newcommand{\ysz}{\ensuremath{y_\mathrm{\mbox{\tiny{SZ}}}}}
\newcommand{\bigysz}{\ensuremath{Y_\mathrm{\mbox{\tiny{SZ}}}}}

\newcommand{\lcdm}{\ensuremath{\Lambda{\rm CDM}}}
\newcommand{\webaddress}{\url{http://pole.uchicago.edu/public/data/sptsz-clusters}}

\defcitealias{vanderlinde10}{V10}
\defcitealias{reichardt13}{R13}

\begin{document}

\title{Galaxy clusters discovered via the Sunyaev-Zel'dovich effect in the 2500-square-degree SPT-SZ survey}
\slugcomment{Accepted by \apjs}

\input{spt_authors}

\begin{abstract}

We present a catalog of galaxy clusters selected via their Sunyaev-Zel'dovich (SZ) effect 
signature from 2500 \degs\ of South Pole Telescope (SPT) data. 
This work represents the complete sample of clusters detected at high significance in the 2500 \degs\ SPT-SZ survey, which was completed in 2011.
A total of  \ncand\ (\ncandfive) cluster candidates are identified above a signal-to-noise threshold of $\xi =4.5$ (5.0). 
Ground- and space-based optical and near-infrared (NIR) imaging confirms overdensities of similarly colored
galaxies in the direction of \nconfirm\ (or $\pctconfirm \%$) of the $\xi>4.5$ candidates
and \nconfirmfive\ (or $\pctconfirmfive \%$) of the $\xi>5$ candidates; the measured purity is consistent with expectations from simulations. 
Of these confirmed clusters, \sptcluster \ were first identified in SPT data, including \newcluster \ new discoveries reported in this work.
We estimate photometric redshifts for all candidates with identified 
optical and/or NIR counterparts; we additionally report redshifts derived from spectroscopic 
observations for \nspec\ of these systems. 
The  mass threshold of the catalog is roughly 
independent of redshift above \mbox{$z \sim 0.25$} leading to a sample of massive clusters that extends to high redshift.  The median mass  of the sample is 
$\mass(\rho_\mathrm{crit}) $ \medianm, the median redshift is 
$z_\mathrm{med}=\medianz$, and the highest-redshift systems are at  $z>$1.4. 
The combination of large redshift extent, clean 
selection, and  high typical mass makes this cluster sample of particular interest for cosmological analyses and 
studies of cluster formation and evolution.  

\end{abstract}


\bigskip\bigskip

\section{Introduction}
\label{sec:intro}

\setcounter{footnote}{0}

Galaxy clusters are the largest collapsed objects in the universe,  
and their abundance is 
exponentially sensitive to the conditions and processes that govern the cosmological growth of structure (see \citealt{allen11} and references therein). 
In particular, measurements of cluster abundance have the power to constrain the 
matter density, the amplitude and shape of the matter power spectrum, and the sum of the neutrino 
masses \citep{wang98,wang05,lesgourgues06b}, and to test models of the cosmic acceleration \citep{wang98,haiman01,weinberg13} in ways that complement constraints from other observational methods.
Clusters are also unique laboratories in which to characterize the interplay between gravitational and astrophysical processes (see \citealt{kravtsov12} for a recent review). 
Well-defined cluster samples over a broad redshift range are critical for such studies. 

Large samples of clusters were first compiled from optical and infrared datasets, in which clusters are 
identified as overdensities of galaxies.
Clusters identified in this manner typically contain of tens to thousands of galaxies (e.g., \citealt{abell58,gladders00,koester07, eisenhardt08,wen12,rykoff14}). 
Clusters of galaxies are also bright sources of extended X-ray emission: the majority of the baryonic matter in clusters (70-95\%) is found in the intracluster medium (ICM) in the form of diffuse gas that has been heated by adiabatic compression and shocks to X-ray emitting temperatures of $10^{7}$ to $10^{8}$ K (see review by \citealt{voit04}). 
The observational expense of detecting high-redshift systems currently limits the size of X-ray samples; however, the tight correlation between ICM observables and the underlying cluster mass enables even modest samples of X-ray-selected systems to place competitive constraints on cosmological models (e.g., \citealt{vikhlinin09,mantz10a}).

Over the past decade, clusters have begun to be identified via their signature in the millimeter-wave sky.
As photons from the cosmic microwave background (CMB) pass through a galaxy cluster, roughly 1\% of the photons will inverse-Compton scatter off the energetic electrons in the ICM. 
This imparts a characteristic spectral distortion to the CMB, known as the thermal Sunyaev-Zel'dovich (SZ) effect \citep{sunyaev72}.
The observed temperature difference, $\Delta T$, relative to the mean CMB temperature, $\Tcmb$, is
\begin{equation}
\label{eq:sze}
\begin{split}
\Delta T &= \Tcmb \ \fsz(x)\int n_\mathrm{e} \frac{k_\mathrm{B}T_\mathrm{e} }{m_\mathrm{e}c^{2}} \sigma_\mathrm{T} dl  \\
  &\equiv  \Tcmb \ \fsz(x) \ \ysz
\end{split}
\end{equation}
where the integral is along the line of sight, \mbox{$x \equiv h\nu$/$k_\mathrm{B} \Tcmb$}, $k_\mathrm{B}$ is the Boltzmann constant, $c$ the speed of light, $n_\mathrm{e}$ the electron density, $T_\mathrm{e}$ the electron temperature, $\sigma_\mathrm{T}$ the Thomson cross-section, 
 $\fsz(x)$ encodes the frequency dependence of the thermal SZ effect:  
\begin{equation}
\fsz(x) = \left ( x\frac{e^{x} + 1}{e^{x}-1} - 4 \right ) (1 + \delta_\mathrm{rc})
\end{equation}
\citep{sunyaev80}, and $\ysz$ is the Compton $y$-parameter.
The term $\delta_\mathrm{rc}$ represents relativistic corrections \citep{nozawa00}, which 
become important at \mbox{$T_\mathrm{e} \gtrsim 8 \ \mathrm{keV}$}.
This frequency dependence leads to a decrement of observed photons (relative to a blackbody at \Tcmb) below the thermal SZ null frequency of \mbox{$\sim$220} GHz and a corresponding increment of photons above this frequency.
The surface brightness of the thermal SZ effect is independent of redshift, 
and the integrated thermal SZ signal is  expected to be a low-scatter proxy for the cluster mass,
as it is proportional to the total thermal energy of the ICM \citep{motl05}.
These properties make cluster samples produced by SZ surveys attractive for cosmological analyses  \citep{carlstrom02}. 

The observed temperature distortions in the CMB caused by the thermal SZ effect are small, typically on the level of hundreds of $\mu \Kcmb$
 for the most massive clusters.\footnote{Throughout this work, noise units and amplitudes expressed in terms of $\Kcmb$ refer to the equivalent deviations in temperature from a 2.73 K blackbody required to produce the observed signals.}
The development over the past decade of high-sensitivity bolometric cameras has enabled the wide and deep surveys required to find these rare systems via the SZ effect.
The first discovery of a previously unknown cluster through its SZ signature was published in 2009 \citep{staniszewski09}; 
today such discoveries have become routine, with catalogs produced by the South Pole Telescope (SPT), Atacama Cosmology Telescope (ACT), and \planck\ teams containing tens to hundreds of massive clusters out to \mbox{$z \sim1.5$} \citep{reichardt13, hasselfield13,planck13-29}.

In this paper we present a cluster catalog extracted 
from the full 2500~\degs \ SPT-SZ survey. 
This sample consists of \ncand  \ galaxy cluster candidates detected at SPT statistical significance $\xi > 4.5$, of which \nconfirm \ (\purityfourfive) have been confirmed as clusters via dedicated optical and near-infrared (NIR) follow-up imaging.  
Follow-up imaging was obtained for all \nfourpseven  \ candidates detected above $\xi=4.7$ and 119 (81\%) of the remaining \nlow \ candidates down to $\xi=4.5$. 
For all confirmed clusters, we report photometric redshifts---or, where available, spectroscopic redshifts---and estimated masses. 
Of the confirmed clusters,  \newcluster \ (\newclusterpercent) are reported for the first time in this work.
Masses are computed using the framework developed for \citet[][hereafter R13]{reichardt13}; we report masses using the best-fit  $\xi$-mass relation  for a fixed flat $\Lambda$CDM cosmology with $\Omega_{\textrm{m}} = 0.3$, $h=0.7$ and $\sigma_{8} = 0.8$. 
 A detailed cosmological analysis incorporating new information from follow-up X-ray observations will be presented in de Haan et al. (in preparation, hereafter referred to as dH14).

This paper is organized as follows.  
Observations and map-making are described in \S \ref{sec:obs}. 
The extraction of galaxy clusters from the maps is detailed in \S \ref{sec:extract}. 
Building off previous work presented in  \citet[][hereafter S12]{Song12b}, we describe our optical and NIR follow-up campaign in \S \ref{sec:followup} and associated confirmation of clusters and measurement of redshifts in \S \ref{sec:redshifts}.
In \S \ref{sec:catalog} we present the full sample of galaxy cluster candidates from the 2500 \degs \ SPT-SZ cluster survey  and highlight particularly notable clusters, and
we conclude in \S \ref{sec:conclusions}. 
All masses are reported in terms of \mass, where \mass \ is defined as the mass enclosed within a radius at which the average density is 500 times the critical density at the cluster redshift.
Selected data reported in this work, as well as future updates to the properties of these clusters,  will be hosted at \webaddress.


\section{Observations and Data Reduction}
\label{sec:obs}
\subsection{Telescope and Observations}
The SPT \citep{carlstrom11} is a 10 m diameter telescope located at the National Science Foundation Amundsen-Scott South Pole station in Antarctica. 
From 2008 to 2011 the telescope was used to conduct  the SPT-SZ survey, a survey of $\sim$2500 \degs \ of the southern sky at 95, 150, and 220 GHz .
The survey covers a contiguous region from 20$^{h}$ to 7$^{h}$ in right ascension (R.A.) and $-65$ to $-40$$^\circ$  \ in declination (see, e.g., Figure 1 in \citealt{story13})  and was mapped to depths of approximately 40, 18, and 70 \muk-arcmin at 95, 150, and 220 GHz respectively. 

The telescope was designed for high-resolution measurements of the CMB, with particular 
attention to the science goal of discovering high-redshift galaxy clusters through the SZ effect.
The large primary mirror leads to beam sizes of roughly 1.6\arcmin, 1.1\arcmin, and 1.0\arcmin \ at 95, 150 and 220 GHz.
Beams of this scale are well matched to the expected angular size of massive clusters at high redshift.
For a non-relativistic thermal SZ spectrum, the centers of the measured 95 and 150 GHz bands are at 97.6 GHz and 152.9 GHz, while the 220 GHz band is approximately at the thermal SZ null.

The 2500 \degs\ SPT-SZ survey was not observed as one contiguous patch; rather it was broken into 19 subregions, or fields, individually scanned to survey depth. 
These fields range in size from $\sim70$ to 250 \degs, and their borders overlap slightly ($\sim4\%$). 
The total area searched for clusters over all fields, 
after correcting for point-source masking (see \S\ref{sec:extract}) 
is \areasearched \ \degs.

The majority of fields were observed in an identical fashion: the telescope was scanned back and forth across the width of the field in azimuth and then stepped in elevation; the scan and step procedure was repeated until the full field had been imaged. 
The process, which constitutes a single \emph{observation} of the field, took between 0.5 and 2.5 hours, depending on the field size and the elevation step. Each
field was imaged in this fashion at least 200 times, and final maps are the sum of all individual observations.
One field, \textsc{ra21hdec$-$50},\footnote{Fields are named via the R.A. and declination of their centers.} was observed with two different strategies: roughly one-third of the data were obtained with the strategy described above, while the remainder
of the observations were conducted by scanning the telescope in elevation at a series of fixed azimuth angles. 
 In Table \ref{tab:fields} we have listed the name, location, area, and depths at 95, 150, and 220~GHz for each field, as well as the year in which the field was observed.
The depth estimates are obtained as in \citet{schaffer11}, using the Gaussian beam approximation 
described in that work.

\begin{figure*}[ht]
\begin{center}
\subfigure[95~GHz minimally filtered map cutout]{\label{fig:map_minfilt_90} 
  \includegraphics[width=2.95in]{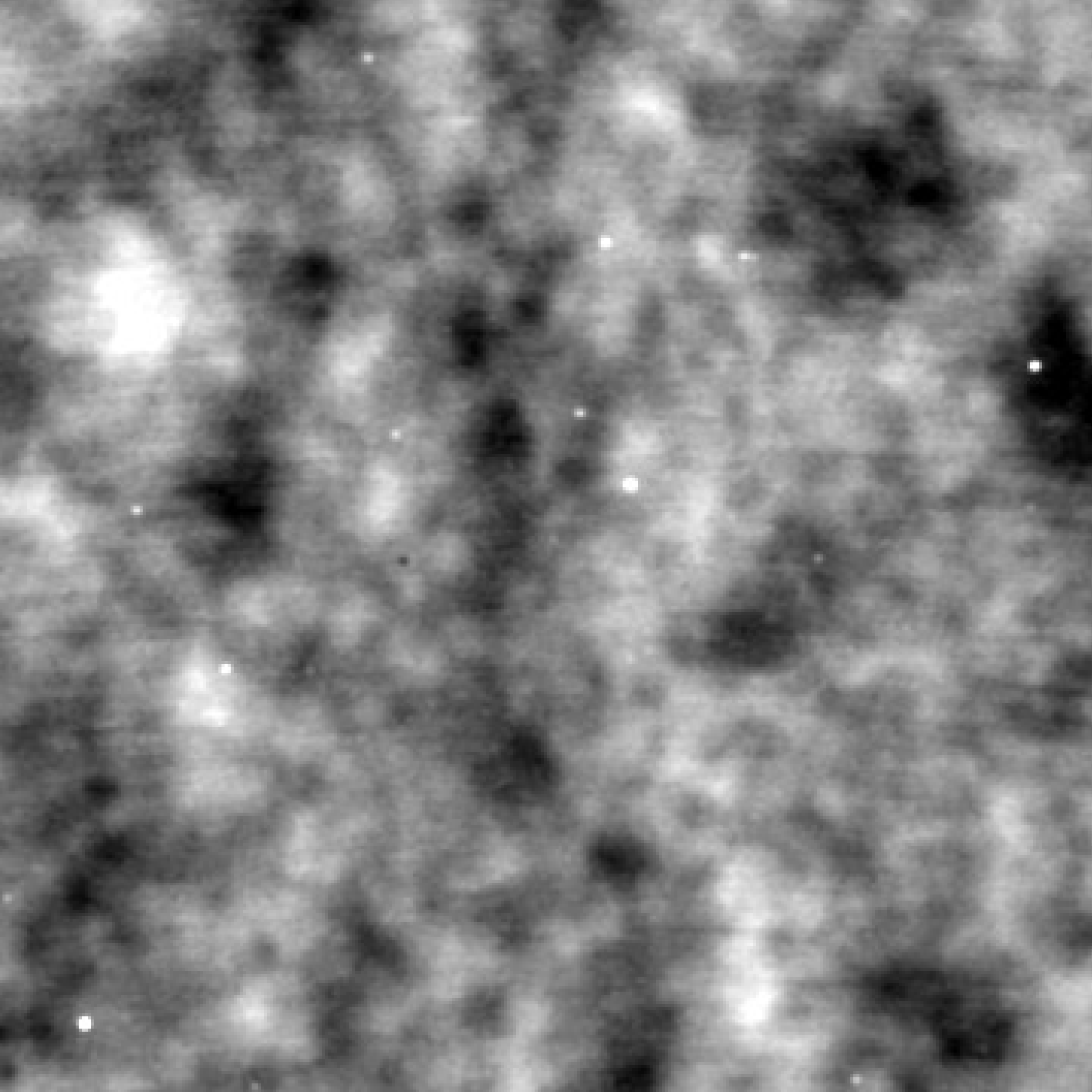}}
\subfigure[150~GHz minimally filtered map cutout]{\label{fig:map_minfilt_150} 
  \includegraphics[width=2.95in]{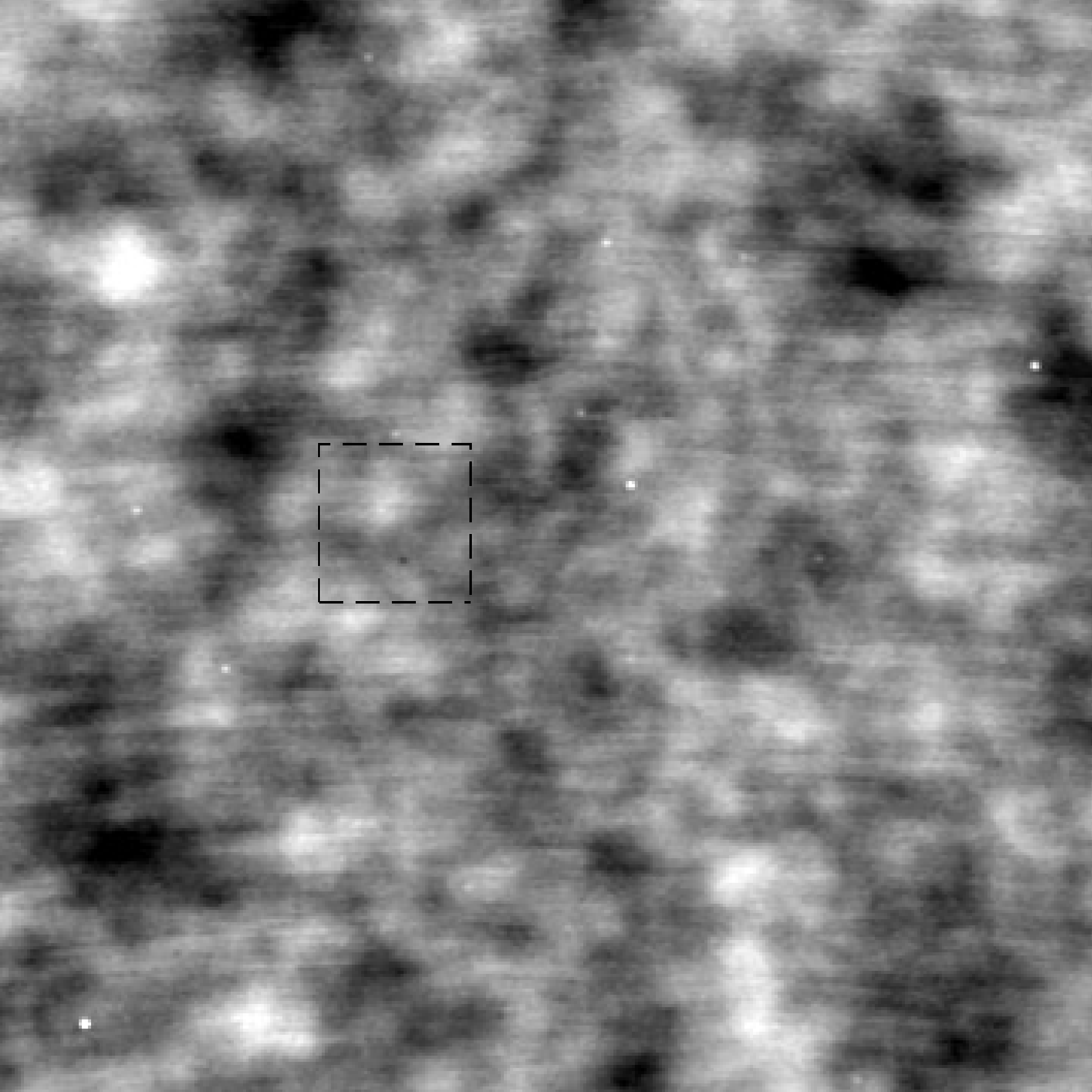}}
\subfigure[Azimuthally averaged cluster-matched two-band filter]{\label{fig:optfilts} 
  \includegraphics[width=2.95in]{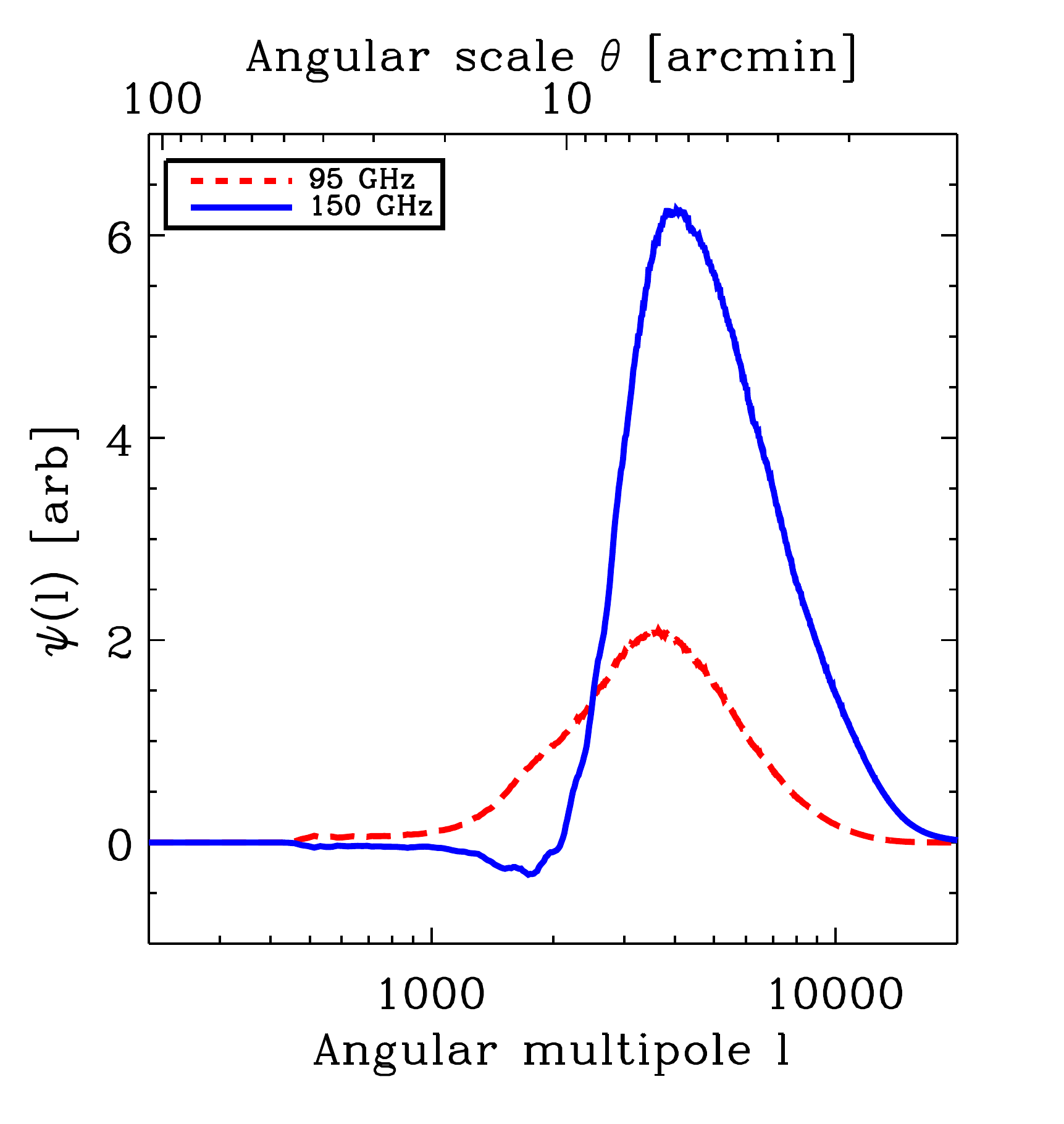}}
\subfigure[Cluster-filtered map, zoomed in to $1^\circ$-by-$1^\circ$]{\label{fig:map_clusfilt_zoom} 
  \includegraphics[width=2.95in]{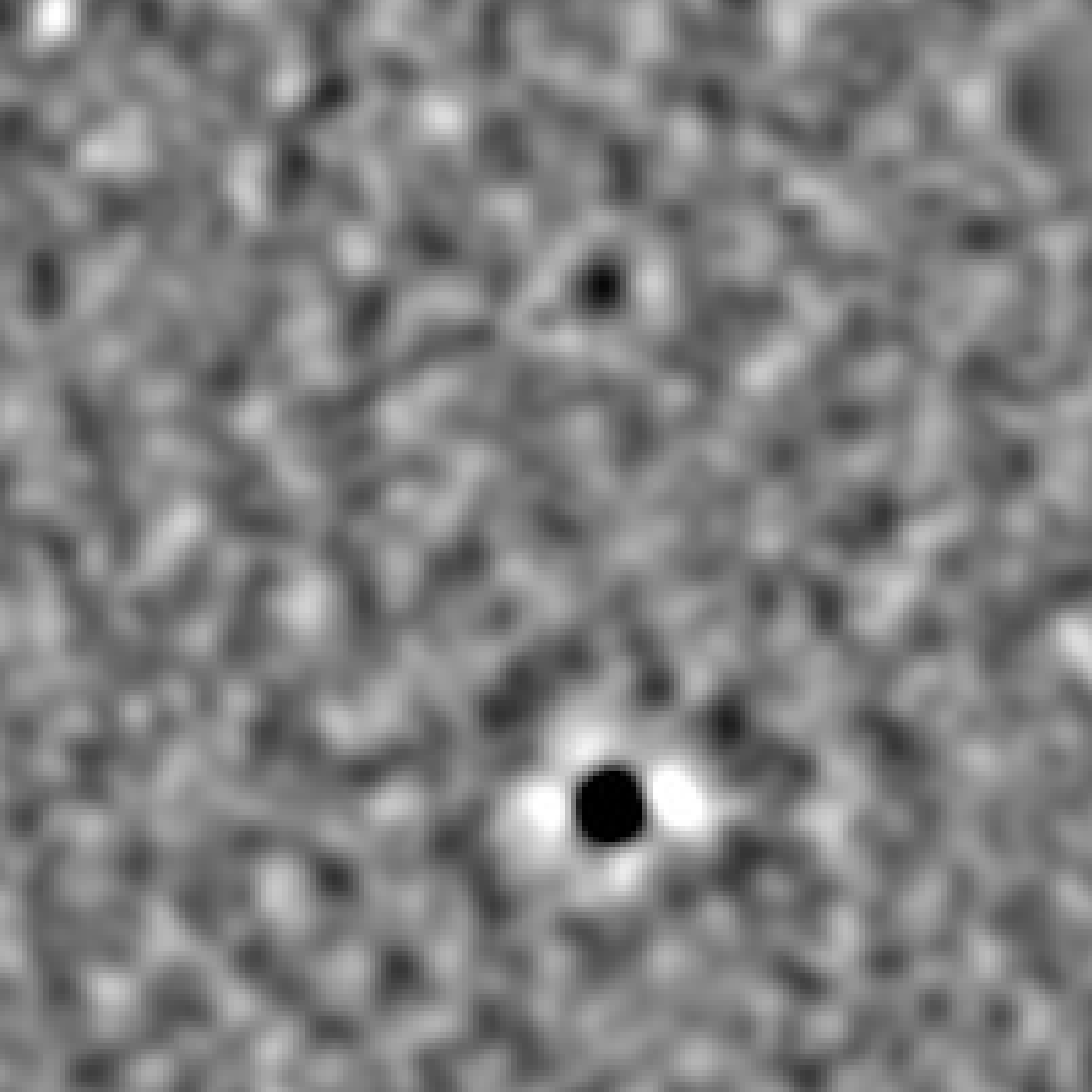}}
\end{center}
\caption{\label{fig:fourpanel} 
Visual representation of the SPT-SZ data and matched filtering process described in 
\S\ref{sec:obs} and \S\ref{sec:extract}. Panels (a) and (b) show $6^\circ$-by-$6^\circ$ cutouts of 
95 and 150~GHz maps from the \textsc{ra21hdec$-$60} field; the displayed temperature range is $\pm 300 \mu K$.  
These maps are made from
data that have been only
minimally filtered (scan-direction high-pass filter at $l$$\sim$50) and 
show the main features of SPT-SZ survey data: large-scale primary CMB fluctuations, emissive
point sources, and SZ decrements from galaxy clusters. 
Panel (c) shows the azimuthally averaged spatial-spectral filter optimized for detection of $\theta_{c} =0\farcm25$ clusters, 
with the red-dashed (blue-solid) curves showing the Fourier-domain coefficients for the 95 (150)~GHz data.
Panel (d) shows a zoomed-in
view of the $1^\circ$-by-$1^\circ$ area delineated by the dashed box in panel (b) after the 
spatial-spectral filter has 
been applied. This map is in units of signal-to-noise, 
and the displayed range is $-5 < S/N < 5$.
Visible in this panel are the $\xi=22.2$, $z=1.132$ cluster SPT-CL~J2106$-$5844
and the $\xi=4.6$, optically unconfirmed candidate SPT-CL~J2106$-$5820. 
}
\end{figure*}

The top two panels of Figure \ref{fig:fourpanel} show $6^\circ$-by-$6^\circ$ cutouts of  
95 and 150~GHz maps from the \textsc{ra21hdec$-$60} field. The maps shown here have 
been very minimally filtered (high-passed at an angular multipole value of
roughly $l=50$ in the scan direction) and show the main signal components present in SPT-SZ survey
data, namely the large-scale primary CMB fluctuations, emissive point sources, and SZ decrements 
from galaxy clusters. The data described in the next section and used as input to the cluster finding
pipeline are more strongly filtered in the time domain before being processed into maps.

\begin{deluxetable*}{ l r c r c c c c}
\tabletypesize{\scriptsize}
\tablecaption{The fields observed by the South Pole Telescope between 2008 and 2011\label{tab:fields}}
\tablewidth{0pt}
\tablehead{
\multicolumn{1}{l}{Name} &
\multicolumn{1}{c}{R.A.}  &
\multicolumn{1}{c}{$\delta$} &
\multicolumn{1}{c}{Area} &
\multicolumn{1}{c}{$\sigma_{95}$} &
\multicolumn{1}{c}{$\sigma_{150}$} &
\multicolumn{1}{c}{$\sigma_{220}$}&
\multicolumn{1}{c}{Survey Year} \\
\colhead{}  &
\colhead{($^\circ$)} &
\colhead{($^\circ$)} &
\colhead{(\degs)} &
\colhead{(\muk-arcmin)} &
\colhead{(\muk-arcmin)} &
\colhead{(\muk-arcmin)} &
\colhead{}
}
\startdata
\input{field_tablevfeb28}
\enddata

\tablecomments{Descriptive information for the 19 fields that comprise the  2500 \degs \ SPT-SZ survey. Here we list the field name, center,  effective area following point source masking, noise levels at 95, 150, and 220 GHz, and the year the field was imaged. The noise levels are estimated as in \citet{schaffer11}, using the Gaussian beam approximation.}
\end{deluxetable*}

\subsection{Map Making}
\label{sec:mapmaking}
A series of operations are performed to convert the raw data from the field observations to calibrated maps. 
The map-making process in this work  is almost identical 
to that in R13; the most significant change lies in the treatment of the 
\textsc{ra5h30dec$-$55} and \textsc{ra23h30dec$-$55} fields.
These two fields, originally observed in 2008, were re-observed in 2010 (\textsc{ra23h30dec$-$55}) or 
2011 (\textsc{ra5h30dec$-$55}) in order to add coverage at 95~GHz (the array fielded in 2008 did
not have enough high-quality 95~GHz detectors to produce survey-depth data). These additional observations
also resulted in deeper final maps at 150~GHz (see Table \ref{tab:fields}).
For these two fields, R13 analyzed the 150~GHz maps described in \citet{vanderlinde10}.
The filtering and calibration for these two fields were treated differently from the other fields. 
In this analysis  we use the full, two-season data (including the 95~GHz data), and the treatment of filtering and calibration
 is uniform across the survey.

We briefly summarize the map-making process here; these steps are described in detail in \citet{schaffer11}.

\begin{itemize}

 \item For each observation, time-ordered detector data are first notch-filtered to
 remove sensitivity to the pulse-tube cooler of the cryostat housing the SPT-SZ receiver. 
 Which detectors' data to include in mapmaking is then
 determined using a series of cuts based upon noise performance and response to
 both a chopped thermal source and on-sky sources. 
 Following
 these cuts the array is ``flat-fielded'' by adjusting each detector's 
 data according to its response relative to other detectors of the same frequency.

 \item 
In every scan across the field, each detector's data is high-pass filtered by removing
the best-fit Legendre polynomial (with the polynomial order depending on the length 
of the scan) and a series of sine and cosine modes. The resulting filter has an effective
cutoff frequency corresponding to 
 angular multipole $l= 400$ (roughly 1/2 degree scales) in the scan direction.
 This filtering step removes
 large-scale noise from the  atmosphere and low-frequency noise from the readout.
 To further reduce atmospheric contamination,
 the mean signal from detectors
 in a single wedge\footnote{The modular SPT-SZ receiver consists of 6 independent
   sub-arrays or ``wedges'' of 160 bolometers that operate at a single
 frequency.} is subtracted from the data of  all detectors in that wedge. 
 This common-mode subtraction
 acts as an isotropic high-pass filter with a cutoff at approximately $l = 500$.
Bright point sources detected at $> 5 \sigma$ ($\sim$ 6 mJy at nominal survey depth) in 150~GHz data are masked
in both of these filtering steps.

 \item Following filtering, the telescope pointing model is used to project the data onto two-dimensional maps. For this analysis we use the 
Sanson-Flamsteed projection \citep{calabretta02} in which pixel rows in the map correspond to constant elevation. 
As (for the majority of observations) the telescope also scans at constant elevation, this projection simplifies the characterization of the applied filtering at the cost
 of slight shape distortions at the map edges.

\item Individual maps, weighted by their noise properties at 1500 $<
  l <$ 4500,  are then coadded to produce the final maps. Maps with anomalously high or low weights or noise
  are not included in the coadd.

\item A calibration factor based on observations of the galactic HII region RCW38  is applied to provide the absolute temperature calibration
  for the maps \citep{staniszewski09}.  
  We have repeated the cluster-finding procedure described in \S \ref{sec:extract} using maps with a CMB-based
  calibration and find negligible differences in the detection significances of galaxy clusters. 
  
\end{itemize}


\section{Cluster Extraction and mm-wave Characterization}
\label{sec:extract}

In this section, we provide a summary of the process by which galaxy cluster candidates are identified and characterized in the SPT  survey data. 
This procedure is almost identical to that used in recent SPT publications; readers are referred to \citet{williamson11} and R13 in particular  for more details. 
The small differences between R13 and this analysis are discussed in \S \ref{sec:compare}.

\subsection{Cluster Extraction}

As described in \S \ref{sec:obs}, the SPT-SZ survey fields are observed at 3 frequency bands centered at approximately 95, 150 and 220 GHz. 
These maps contain signal from a range of astrophysical sources. 
For the purposes of this analysis, we characterize the observed temperature, $T$, in the maps at frequency $\nu_i$ and location {\bf x} by:
\begin{equation}
\begin{split}
T(\mathbf{x},\nu_i) =  B(\mathbf{x},\nu_i) * [ \fsz(\nu_i) \Tcmb \ysz(\mathbf{x}) + 
n_\mathrm{astro}(\mathbf{x},\nu_i)]   \\
+ n_\mathrm{noise}(\mathbf{x},\nu_i). 
\end{split}
\end{equation}
Here $B$ encompasses the effects of the beam and applied
filtering;
the expected thermal SZ signal
is given by the product of the frequency dependent term $\fsz$,
the CMB temperature 
\Tcmb, and the Compton-$y$ parameter \ysz; $n_\mathrm{astro}$
encompasses astrophysical signals---all of which are modeled here as Gaussian noise---and $n_\mathrm{noise}$ corresponds to instrumental and
residual atmospheric noise not removed by the filtering discussed in
\S \ref{sec:obs}.  For SPT maps, $n_\mathrm{astro}$ primarily consists of lensed primary CMB fluctuations,
kinetic and thermal SZ from the
clusters below the SPT detection threshold, and dusty extragalactic sources; radio sources below the SPT detection
threshold contribute negligibly to the maps.
As in previous
work, we model these
noise terms based upon recent SPT power spectrum constraints
\citep{keisler11,shirokoff11}.  

Given the known spatial and spectral characteristics of galaxy clusters as well as the sources of noise in the maps, 
we construct a filter
designed to maximize our sensitivity to galaxy clusters \citep{melin06}. 
This Fourier-domain filter takes the form:
\begin{equation}
\boldsymbol{\psi}(\mathbf{l},\nu_i) = \sigma_\psi^{-2} \ 
\sum_j \mathbf{N}_{ij}^{-1}(\mathbf{l}) \fsz(\nu_j) S_\mathrm{filt}(\mathbf{l},\nu_j),
\label{eqn:optfilt}
\end{equation}
where the predicted variance in the filtered map, $\sigma_\psi^{-2}$, is given by
\begin{equation}
\begin{split}
\sigma_\psi^{-2} = \int d^2 l \sum_{i,j} \fsz(\nu_i) S_\mathrm{filt}(\mathbf{l},\nu_i) \ \mathbf{N}_{ij}^{-1}(\mathbf{l}) \ \times \\
  \fsz(\nu_j) S_\mathrm{filt}(\mathbf{l},\nu_j),
\end{split}
\end{equation}
$\mathbf{N}$ is the Fourier-domain version of the band-band, pixel-pixel covariance matrix, and
$S_\mathrm{filt}$ is the Fourier transform of  the real space cluster template convolved with $B(\mathbf{x},\nu_i)$.  
We use a projected isothermal $\beta$-model \citep{cavaliere76} with $\beta$ fixed to 1 as our source template:
\begin{equation}
\Delta T = \Delta T_{0}(1+ \theta^{2}/\theta_\mathrm{c}^{2})^{-1}
\label{eqn:beta}
\end{equation}
where $\Delta T_{0}$ is the normalization, $\theta$ is the angular separation from the cluster center, and $\theta_\mathrm{c}$ is the core radius. 
As discussed in \citet{vanderlinde10}, given the spatial resolution of  the SPT this simple profile is adequate for our purposes: no improvement in the detection of clusters is seen using   
more sophisticated models (e.g., \citealt{nagai07}, \citealt{arnaud10}). 

A series of profiles with evenly-spaced core radii ranging from $0\farcm25$  \ to 3\arcmin \ is used to construct 12 matched filters. 
Azimuthal averages of the $0\farcm25$ matched-filter coefficients for 95 and 150~GHz are
shown as a function of $\ell$ in the bottom left panel of Figure \ref{fig:fourpanel}.
After the application of these filters to the 95 and 150 GHz maps (the relative noise levels in the 220 GHz maps are too high to significantly improve cluster detection, so these data are omitted here), cluster candidates are extracted via a peak detection algorithm similar to the {\tt SExtractor} routine \citep{bertin96}. 
We record the location and maximum detection significance across all filter scales, $\xi$, for all cluster candidates with $\xi \ge 4.5$. 
The bottom-right panel of Figure \ref{fig:fourpanel} shows the result of applying the $0\farcm25$  \
matched filter to 95 and 150~GHz maps of the \textsc{ra21hdec$-$60} field. 

We take two steps to reduce the number of spurious sources created by filter artifacts (in particular decrements created by the filter ``ringing" around strong sources). 
First, prior to filtering the maps we mask a 4\arcmin \ region around all point sources detected above 5$\sigma$ in 150 GHz maps optimized for point source detection. 
Following filtering we additionally exclude cluster candidates detected within 8\arcmin \ of these emissive sources. 
This step removes \areamasked \ \degs \ from the survey. 
We perform a separate cluster-finding analysis on these masked regions and report on those detections
(which we do not include in the official catalog) in \S\ref{sec:clustersinmask}.
We also find it necessary to veto candidates around the strongest cluster detections: we  exclude candidates within a 10\arcmin \ region around  $\xi>20$ detections.  
This final cut removes less than 1 \degs \ from the entire survey. 

\subsection{Integrated Comptonization}
\label{sec:ysz}
For every cluster candidate  we measure the integrated Comptonization, \bigysz, within a $0\farcm75$ radius aperture.   
For a projected isothermal $\beta$-model, \bigysz\ is defined as:
\begin{equation}
\bigysz^{0\farcm75} = 2\pi \int_0^{0\farcm75} y_{0} (1+ \theta^{2}/\theta_\mathrm{c}^{2})^{-1} \theta d\theta
\label{eqn:bigysz}
\end{equation}
where $y_{0}$ is the peak Comptonization. 
The aperture  is slightly smaller than  the 1\arcmin \ radius utilized in R13: 
as demonstrated in \citet{saliwanchik15},
at the resolution and noise levels of the SPT-SZ survey, 
the scatter of \bigysz \ at fixed cluster mass is minimized at $\theta=0\farcm75$. 
This scatter, measured to be  $27 \pm 2$\%, is comparable to the scatter in the relation between SPT detection significance and mass that is used to estimate cluster masses in R13 and other SPT cluster publications.

\bigysz\ is computed using the same procedure as in R13. 
Briefly, for every cluster candidate, the likelihood of the observed two-band SZ signal given the model of 
Eqn.~\ref{eqn:beta} is estimated using a simple gridded parameter search. For every point in the four-parameter 
space of $y_0$, $\theta_\mathrm{c}$, $x$ position, and $y$ position, the value of $\bigysz^{0\farcm75}$ is 
calculated using Eqn.~\ref{eqn:bigysz}, and the best-fit value and $1 \sigma$ constraints are estimated 
from the (one-dimensional) $\bigysz^{0\farcm75}$ posterior distribution. Effective step-function priors
are placed on the values of $\theta_\mathrm{c}$, $x$, and $y$ by restricting the parameter grid for each cluster 
such that $x$ and $y$ are within $1\farcm5$ of the best matched-filter position for that cluster 
(we note that the absolute difference between the best-fit and matched-filter positions is $<0\farcm75$ for $97\%$ of the
cluster candidates) and 
that the physical core radius of each cluster is $50 \ \mathrm{kpc} \le r_\mathrm{core} \le 1 \ \mathrm{Mpc}$.
For unconfirmed candidates we use a redshift of $z=1.5$ to set the physical scale. 
The priors on $r_\mathrm{core}$ are motivated by the known mass distribution of SPT-selected clusters
and the mass-concentration relation measured by, e.g., \citet{mandelbaum08}.

\subsection{Contamination}
 \label{sec:simulations}
 Simulations are used to estimate the level of contamination in the cluster catalog by false detections from instrumental noise and non-cluster astrophysical signals. 
The simulations are similar to those used in R13; we provide a brief summary here.

 For each of the 19 SPT fields we create 100 simulated sky maps composed of contributions from the CMB, emissive sources, and noise (note that we do not include a thermal SZ contribution when quantifying the expected number of false detections).
The CMB component is modeled as a Gaussian random field based on the best-fit WMAP7 + SPT lensed $\Lambda$CDM  model \citep{komatsu11, keisler11}, and the point source model contains contributions from radio sources  and dusty star-forming galaxies (DSFG): all source contributions are modeled as  Gaussian fields.  
 The simulated radio population follows the results of \citet{dezotti05}, \citet{vieira10}, and \citet{reichardt12b}. We assume 100\% correlation between the bands, a spectral index of $\alpha = -0.53$ and, at 150 GHz, an amplitude of $D_l = l (l +1) C_l$/2$\pi$ = 1.28 \muksq \ at $l=3000$. 
The amplitudes and spectral indices of the DSFG contributions are also constrained by recent SPT measurements \citep{reichardt12b}. At 150 GHz and  $l=3000$ the Poisson contribution has amplitude $D_l$ = 7.54 \muksq \  and the clustered contribution $D_l$ = 6.25 $\muksq$; we use $\alpha=3.6$ for both contributions. 
 For each simulated map, we create noise realizations using jackknife noise maps (e.g., \citealt{sayers09}).
 The noise maps are created by randomly multiplying half of the observations of a field by $-1$ and then coadding the entire set of observations.
 This is a change from the R13 simulations which assumed  stationary Gaussian noise.

\begin{figure}
\includegraphics[width=3.5in]{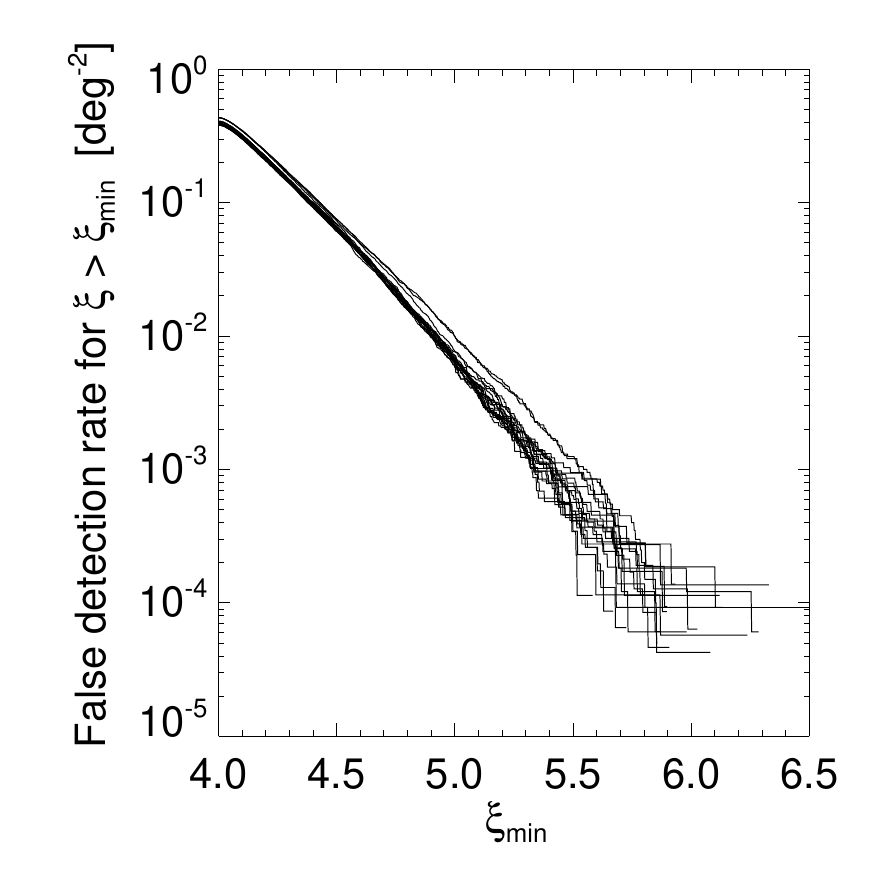}
\caption{Simulated false-detection rate for each of the 19 fields in the SPT-SZ survey.  Plotted is the cumulative density of  false detections above a detection significance, $\xi_\textrm{min}$. 
We expect \nfalsefive \ false detections at $\xi \ge 5$  and \nfalsefourf \ at $\xi \ge 4.5$ for the entire survey.  
\\
}
\label{fig:falsedetection}
\end{figure}

To estimate the expected number of false detections, the cluster-detection algorithm is run  (with point source masks and apodization matching the real data) on these cluster-free simulated maps.  In Figure \ref{fig:falsedetection} we plot the expected rate of false detections in each individual SPT field.  Two fields have slightly higher false-detection rates owing to the inclusion of boundary regions with uneven coverage in the area searched for clusters. In total, across all fields, we expect 172 false detections at $\xi > 4.5$ and 18.5 at $\xi >5$. We return to the question of false detections again in \S \ref{sec:catalog} where we compare our expectations to the measured purity of the cluster sample.

\section{Follow-up Observations}
\label{sec:followup}

We use optical and in some cases NIR imaging to confirm candidates as clusters  and to obtain redshifts for confirmed systems.
In this section, we briefly describe our follow-up strategy and the reduction of imaging data.
Follow-up imaging was obtained for all 530  candidates detected above $\xi=4.7$ and 119 of the remaining 147 \ candidates down to $\xi=4.5$; the remaining 28 low-significance candidates were not imaged owing to limited follow-up resources. 
The procedures discussed here closely follow those presented in S12.

\subsection{Follow-up Strategy}
\label{sec:strategy}
Our candidate follow-up strategy evolved over the course of the SPT-SZ survey.  
Initially, clusters were confirmed using preexisting imaging from the Blanco Cosmology Survey (BCS; \citealt{menanteau10,desai12,bleem15a}, see \S \ref{sec:otherdata}) as well as targeted imaging on the  Blanco/MOSAIC-II\footnote{http://www.ctio.noao.edu/mosaic/} and Magellan/IMACS imagers \citep{dressler06} (see \citealt{high10} for more details). 
As the candidate sample size grew, we adjusted our imaging strategy to effectively incorporate a range of small- and large-aperture telescopes.  
Our resulting strategy is as follows:

\begin{itemize}
\item All cluster candidates are ``pre-screened'' using imaging data from the Digitized Sky Survey (DSS)\footnote{The DSS is a digitization of the photographic sky survey conducted with the Palomar and UK Schmidt 
telescopes; http://archive.stsci.edu/dss/.} to determine if a cluster candidate lies at low redshift. 
We find the DSS images are generally sufficient to allow identification of the optical counterparts of SPT clusters out to redshift $z=0.5$, with a tail extending to $z=0.7$. 
Roughly 50\% of confirmed SPT clusters are identifiable in DSS data; candidates not apparent in the DSS are classified as high-redshift targets.

\item We observe potential low-redshift candidates at 1~m -- 2~m class facilities. 
If the candidate is not confirmed in these data, it is reclassified as a potential high-redshift system. 

\item High-redshift targets are imaged on larger aperture (4~m -- 6.5~m) telescopes.
The majority of these cluster candidates with  $\xi \ge 4.8$ and many candidates with $4.5 \le \xi  < 4.8$ have also been imaged
in the NIR from ground- or space-based facilities. 
This threshold was  $\xi = 4.5$ in S12; the increased threshold here is due to limited NIR resources. 

\item Following the release of the Wide-field Infrared Survey Explorer  (WISE) all-sky catalogs \citep{wright10}, observations of  $\xi < 4.8$ candidates were weighted towards those candidates with lower values of the NIR ``blank-field" statistic presented in S12. 
This statistic quantifies the significance of any overdensity of galaxies at the candidate's location compared to random fields; lower values correspond to more dense regions.
As such,  the follow-up of these lower significance candidates is biased to maximize the number of confirmed clusters.

\end{itemize}

In Table \ref{tab:instruments} we list the facilities and instruments used in our follow-up campaign. 
We assign each telescope/instrument combination a numerical alias which we use in Table \ref{tab:catalog} to identify the source of redshift information. 

\begin{deluxetable*}{llllllcl}
\tabletypesize{\scriptsize}
\tablecaption{Optical and infrared imagers\label{tab:instruments}}
\tablewidth{0pt}
\tablehead{
\colhead{Ref.\tablenotemark{a}} &
\colhead{Site} &
\colhead{Telescope} &
\colhead{Aperture} &
\colhead{Camera} &
\colhead{Filters\tablenotemark{b}} &
\colhead{Field} \\
\colhead{~} &
\colhead{~} &
\colhead{~} &
\colhead{(m)} &
\colhead{~} &
\colhead{~} &
\colhead{~} &
}
\startdata
1 & Cerro Tololo & Blanco & 4 & MOSAIC-II & $griz$ & $36\arcmin \times 36\arcmin$ \\
2 & Las Campanas & Magellan/Baade & 6.5 & IMACS f/2 & $griz$ & $27\farcm4 \times 27\farcm4$ \\
3\tablenotemark{c} & Las Campanas & Magellan/Clay & 6.5 & LDSS3 &$griz$ & $8\farcm3$ diam.\ circle  \\
4\tablenotemark{d} & Las Campanas & Magellan/Clay & 6.5 & Megacam & $gri$  & $25\arcmin \times 25\arcmin$ \\
5 & Las Campanas & Swope & 1 & SITe3 & $BVRI$ & $14\farcm8 \times 22\farcm8$ \\
6 & La Silla & MPG/ESO & 2.2 & WFI & $BVRI$ & $34\arcmin \times 33\arcmin$ \\
7 & La Silla & New Technology Telescope & 3.6 & EFOSC2 & $griz$ & $4\farcm1 \times 4\farcm1$ \\
8 & Cerro Tololo & Blanco & 4 & NEWFIRM & $K_s$ & $28\arcmin \times 28\arcmin$ \\
9 & Las Campanas & Magellan/Baade & 6.5 & FourStar & $J,H,K_{s}$ & $10\farcm8 \times 10\farcm8$ \\
10 & Satellite & Spitzer Space Telescope & 0.85 & IRAC & 3.6$\mu$m, 4.5$\mu$m & $5\farcm2 \times 5\farcm2$ \\
\nodata & Satellite & Wide-field Infrared Survey Explorer & 0.40 &  \nodata &  $W1, W2$ & $47\arcmin \times 47\arcmin$ \\

\enddata
\tablecomments{Optical and infrared cameras used in SPT follow-up observations. }
\tablenotetext{a}{Shorthand alias used in Table \ref{tab:catalog}.}
\tablenotetext{b}{Not all filters were used to image every cluster.}
\tablenotetext{c}{http://www.lco.cl/telescopes-information/magellan/instruments/ldss-3}
\tablenotetext{d}{Megacam data were acquired for a large follow-up weak-lensing program.}
\end{deluxetable*}

\subsection{Targeted Observations and Data Reduction}
\subsubsection{Optical Data}
Our strategy for targeted optical follow-up varied with the aperture of the follow-up telescope. 
In this section, we first describe our
strategy for observing candidates we expect to be at low redshift with 1~m -- 2~m class telescopes; we then
move on to our strategy for observing likely high-redshift candidates with larger telescopes.
Note that in this section we quote depths relative to the apparent magnitude of L$^*$ galaxies: our model for the redshift evolution of these galaxies is described in \S \ref{sec:redshifts}.

In order to rapidly image a large number of systems, we adopt a minimalist approach  for low-redshift candidates observed on the 1 m Swope telescope.  
Based on an initial ``by-eye'' redshift estimate from the DSS screening step\footnote{
These crude redshifts are based on a combination of the color, brightness, and angular size of identified cluster galaxies.
The uncertainty on these estimates is  \mbox{$\sigma_{z}\sim0.1$}, with large outliers owing to variable DSS image quality.}, we choose a pair of filters (\emph{BV,VR,RI}) expected to straddle the 
4000~\AA \ break. 
Three filters are used when required to avoid redshift degeneracies. 
Candidates are imaged to depths sufficient for robust estimation of red-sequence redshifts: we require detection of $0.4$L$^*$ red-sequence galaxies at 8$\sigma$ in the redder filter and 5$\sigma$ in the bluer filter.
A second round of deeper imaging is obtained for systems with significantly underestimated DSS redshifts. 
Non-confirmed candidates and clusters at higher redshift ($z>0.7$) are re-observed on larger class telescopes and/or with NIR imagers.  

Low-redshift candidates are also observed on the MPG/ESO 2.2~m telescope using the Wield Field Imager (WFI; \citealt{baade99}).
These observations are deeper than those acquired on Swope as they were designed to also enable studies of the galaxy populations of these clusters (e.g., \citealt{zenteno11}).
Based  on its preliminary DSS redshift, each candidate is imaged in three filters (\emph{BVR} or \emph{VRI}), to depths sufficient to detect $0.4$L$^*$  galaxies at $10 \sigma$ in the bands straddling the 4000~\AA \ break. The third band is used for photometric calibration (see below). 
Second-pass imaging is obtained as necessary to adjust for imperfect initial redshift estimates. 
As with our Swope program, non-confirmed candidates and clusters at higher redshift ($z>0.75$) are re-observed on larger class telescopes and/or with NIR imagers.

As described in S12, we also adopt a  two-pass strategy for observations on 4~m--6.5~m class telescopes. 
Candidates are first imaged in the \emph{g-}, \emph{r-},  \emph{z-}bands (or \emph{g-}, \emph{r-} and \emph{i-}bands early in the follow-up campaign) to depths sufficient to 
detect $0.4$L$^*$ galaxies at $z = 0.75$ at $5 \sigma$  in the redder bands. 
The \emph{g-}band data is used for photometric calibration. 
Following this first-pass imaging, all non-confirmed candidates at $\xi \ge 4.8$ (and a subsample of non-confirmed candidates below this threshold)
were further imaged in the \emph{z-} and \emph{i-} (or \emph{r-}) bands to extend this redshift range to $z=0.9$ and/or were imaged in the NIR, as described below. 

With the exception of images from Magellan/Megacam \citep{mcleod06}, all optical images were reduced with the PHOTPIPE pipeline \citep{rest05a, garg07,miknaitis07}. 
Megacam images were reduced using the Smithsonian Astrophysical Observatory Megacam reduction pipeline. 
The PHOTPIPE reduction process as applied to SPT clusters is explained in \citet{high10} and the Megacam pipeline in  \citet{high12}. 
All reductions include the standard CCD image processing steps of masking bad or saturated pixels, applying crosstalk and overscan corrections, debiasing, flat-fielding, correcting for scattered light via illumination corrections, and---where necessary in the redder bands---defringing.  
Cosmic rays are also removed from the Megacam images.  
Images are coadded using the {\tt SWarp} algorithm \citep{bertin02}, and astrometry is tied to the Two Micron All Sky Survey (2MASS) catalog \citep{skrutskie06}. 

Sources are detected using the {\tt SExtractor} algorithm (v 2.8.6) in dual-image mode; we use the deepest images with respect to red-sequence galaxies as the detection images. 
As in previous works, photometry is calibrated using Stellar Locus Regression (SLR, \citealt{high09}) with absolute calibration derived using stars in the 2MASS catalog. 

There are a few modifications to the calibration process discussed in previous SPT publications. 
For clusters only imaged in two bands at Swope, we combine \emph{J-}band data from 2MASS with the optical data to create the two colors required for calibration via SLR. 
The SLR calibration for candidates imaged on the NTT was somewhat challenging owing to the small field of view  ($4\farcm1 \times 4\farcm1$) of the EFOSC2 imager \citep{buzzoni84}. 
These calibrations were performed with fewer stars than the other imaging; for a few fields we jointly calibrated the data with other observations from the same night. 
We increase the expected uncertainty on the color calibration for these systems to 5\% (compared to the typical 2-3\% observed with SLR, see \citealt{high09,bleem15a}) and include this extra scatter in our estimates of redshift uncertainties. 
Finally, for a small subset of clusters located in the wings of the Large Magellanic Cloud (LMC), we restrict our fitting to stars with counterparts in the 2MASS catalog to enable convergence of the regression algorithm.\footnote{The stellar envelope of the LMC extends well beyond regions of significant thermal dust emission; we see no evidence of contamination in the SPT mm-wave maps.}
 
\subsubsection{Near-Infrared Data}

 \spitzer /IRAC imaging \citep{fazio04} at 3.6~\um \ and 4.5~\um \ is obtained for the majority of high-redshift SPT cluster candidates at $\xi \ge 4.8$ and a subsample of systems at $4.5 \le \xi \le 4.8$.
In total 276 candidates (241 candidates at $\xi > 4.8$) were observed as part of our \spitzer \ follow-up program.\footnote{Archival observations with varying wavelength coverage and exposure times are available for an additional 16 (typically low-redshift) SPT systems.} 
These \spitzer \ data play a crucial role in confirming and determining the redshift of clusters at $z>0.8$. 
Candidates are imaged in 8 $\times$ 100~s and 6 $\times$ 30~s dithered exposures at
3.6 and 4.5~\um, respectively; the resulting coadded images are  sufficient for the detection  in the 3.6~\um  \ band of $z=1.5$ $0.4$L$^*$ galaxies at $10 \sigma$. 
These observations are reduced following the methodology of \citet{ashby09}. 
Briefly, these reductions correct for column pulldown, mosaic the individual exposures, resample the images to 0\farcs86 pixels (half the solid angle of the native IRAC pixels), and reject cosmic rays. 

Ground-based NIR imaging of some candidates was acquired with the NEWFIRM imager  \citep{autry03} 
at the CTIO $4\,\mathrm{m}$ Blanco telescope and the FourStar imager \citep{persson13} on the Magellan Baade $6.5\,\mathrm{m}$ telescope.  
NEWFIRM data with a target 10$\sigma$ point source depth of 19 Vega magnitudes in the $K_{s}$ filter were obtained for 31 candidates during two runs in November 2010 and July 2011 under photometric conditions. Typical observations consisted of 16 point dither patterns, with 6 $\times$ 10~s exposures obtained at each dither position.
The data were reduced using the \texttt{FATBOY} pipeline, originally developed for the FLAMINGOS-2 instrument, and modified to work with
NEWFIRM data in support of the Infrared Bootes Imaging Survey
\citep{gonzalez10}.  Images were coadded using 
{\tt SCAMP} and {\tt SWarp} \citep{bertin02} and photometry was calibrated to
2MASS.

Additional $JHK_{s}$-band imaging was collected with FourStar for 34 candidates 
during several runs in 2012, 2013, and 2014 in average to good conditions. 
Several exposures were 
taken at 9--15 different pointed positions with the coordinates of the cluster centered on either the mosaic or one of the four detectors. The images were flat-fielded using standard IRAF routines; WCS registering and stacking were done using 
the PHOTPIPE pipeline and were calibrated photometrically to 2MASS.

\subsubsection{Spectroscopic Observations}
\label{sec:specsample}
We have also used a variety of facilities to obtain spectroscopic observations of SPT clusters. 
These observations fall into two categories: small, few-night programs focused primarily on the highest-redshift subset of the SPT-selected clusters (e.g., \citealt{brodwin10, foley11, stalder13,bayliss14}) and longer, multi-semester campaigns. 
The longer programs include observations of high-redshift systems using FORS2 on the VLT \citep{appenzeller98} and a large survey program on the Gemini-South telescope (NOAO PID 2011A-0034) using GMOS-S \citep{hook04} that targeted 85 SPT clusters in the redshift interval $0.3 < z < 0.8$. In \citet{ruel14} we describe in detail our spectroscopic followup campaign and report spectroscopic redshift measurements for 61 SPT clusters and velocity dispersions for 48 of these clusters. 
Here we report an additional \nspecnew \ cluster redshifts using newly obtained data from the Gemini Survey Program. 
As described below in \S \ref{sec:catalog}, we also search the literature for spectroscopic counterparts of SPT clusters; in total \nspec \ of the clusters in this work have spectroscopically measured redshifts.

\subsection{Other Datasets}
\label{sec:otherdata}
In addition to dedicated optical/NIR observations of SPT cluster candidates, we use imaging from three surveys that overlap the SPT footprint: the BCS, the \spitzer-South Pole Telescope Deep Field (SSDF; \citealt{ashby13}), and the WISE all-sky survey. 
The BCS is a $\sim$80 \degs \ 4-band  (\emph{g,r,i,z}) survey (NOAO large survey program 2005B-0043) with imaging sufficient for cluster confirmation to \mbox{$z\sim$1}.
It is composed of two fields roughly centered at (R.A.,DEC) = (23h,$-55$d) and (5h30m,$-53$d). These fields roughly 
overlap with the \textsc{ra5h30dec$-$55} and \textsc{ra23h30dec$-$55} fields, the first fields surveyed by the SPT (See Table \ref{tab:fields}).  
We use the reductions presented in \citet{bleem15a} in this work. 

The SSDF, a  94 \degs \ survey at  3.6 and 4.5 $\mu$m, is centered at (R.A.,DEC)= (23h30,$-55$d). 
It encompasses a large fraction of the \textsc{ra23h30dec$-$55} field. 
This survey, one of the largest extragalactic surveys ever conducted with \spitzer/IRAC, has imaging sufficient for cluster confirmation and redshift estimation to \mbox{$z\gtrsim1.5$}. 

The WISE all-sky survey provides catalogs and images of the entire sky in the W1-W4 bands (3.4--22~\um)\footnote{http://wise2.ipac.caltech.edu/docs/release/allsky/};
the shorter-wavelength NIR data from WISE are sensitive to cluster galaxies out to  \mbox{$z\sim 1.3$} \citep{gettings12,stanford14}. 
As discussed in \S \ref{sec:strategy}, we use WISE data to prioritize the follow-up of lower significance candidates.

\section{Cluster Confirmation \&  Redshift Estimation}
\label{sec:redshifts}

As in previous SPT publications, we deem a candidate to be ``confirmed'' if we identify an excess of clustered red-sequence galaxies at the SPT location.
In this section, we describe our model for the optical and NIR properties of red-sequence galaxies, the process by which we identify excesses of such galaxies at candidate locations, and the estimation of redshifts for confirmed clusters using optical and/or NIR data. 
Finally, for unconfirmed candidates, we describe our procedure for determining the redshift to which our imaging is sufficient to confirm the candidate as a cluster.

\subsection{Red-Sequence Model}
We create our model for the color-magnitude relation of red-sequence galaxies using the {\tt GALAXEV} package \citep{bruzual03}.
We model the galaxies as passively evolving, instantaneous-burst stellar populations with a formation redshift of $z = 3$; 
the stellar populations are generated using  the Salpeter initial mass function \citep{salpeter55} and follow the Padova 1994 evolutionary tracks \citep{fagotto94}.
Metallicites are chosen based upon analytical fits to RCS2 cluster
data \citetext{Koester, private comm.}, and cubic splines are used to interpolate the discrete output of the code to arbitrary redshifts.
We compare our stellar-synthesis $m^{*}(z)$ model to the \citet{rykoff12} model for the maxBCG cluster sample over the redshift range for which that model is valid ($0.05 < z < 0.35$). 
We find a small ($\sim$0.2 mag) offset, and we
correct our model for this offset.
Our red-sequence model is further calibrated to real data using the SPT spectroscopic  subsample as described below. 

\subsection{Identifying Red-Sequence Overdensities and Estimating Optical Redshifts}

For each cluster candidate, we search for a redshift at which there is a clear 
excess of galaxies near the 
candidate position that are consistent with the expected red sequence at that redshift.
We typically search for overdensities out to \mbox{$z \sim 0.7$} (\mbox{$z \sim1$}) for systems with first-pass  (second-pass) optical imaging.
At a series of discrete redshifts in this range, we
compute the background-subtracted, weighted red-sequence galaxy count in a 2\arcmin \ (3\arcmin \ at $z<0.3$) 
region around the SPT position.  
The contribution of each galaxy to the weighted sum is based upon the consistency of the galaxy's color and magnitude
with the red-sequence model  at the redshift in question. 

To confirm a cluster, we require a significant peak in background-subtracted weighted counts.   
The preliminary redshift is identified as the location of this peak. 
In a few instances where this peak is marginally significant (for example, when the cluster is well-detected in one imaging band but poorly in the second owing to incomplete follow-up),  we confirm clusters based solely on visual identification of member galaxies and manually select cluster galaxies for redshift estimation. 

To further refine the preliminary redshift and to estimate a statistical uncertainty,
we next  bootstrap resample the galaxies that contribute to the peak.
 For each bootstrap sample, the estimated redshift is the
redshift at which the $\chi^{2}$ statistic:
\begin{equation}
\chi^{2} = \sum\limits_\mathrm{galaxies} \frac{[\mathrm{Model(magnitude, \
  color, \ z)} -
  {\bf g} ]^{2}}{\mathrm{color \ error}^{2} + \sigma_\textrm{rs}\ ^{2}}
\end{equation} 
is minimized.
Here {\bf g} encodes the color and magnitude of the galaxies, and
\emph{$\sigma_\textrm{rs}$} = 0.05 \citep{koester07,mei09} is the intrinsic
spread of the red sequence.  
The reported redshift is the median redshift of 100 bootstrap resamples.  
The final redshift uncertainties are reported as the statistical error estimated from this bootstrap resampling process (typically small as most estimates are derived from tens of galaxies) added in quadrature with a redshift-dependent scatter determined during the spectroscopic tuning of the red-sequence model, which we now describe.

We first estimate ``raw" redshifts using the uncalibrated red-sequence model for 103 clusters with good follow-up data and spectroscopic redshifts in the SPT sample. 
As follow-up observations span different instruments with different combinations of filters  (e.g., subsets of \emph{griz} on IMACS, LDSS3, Megacam, MOSAIC-II and EFOSC2 and \emph{BVRI} on the SITe3 and WFI), we separately calibrate models for each color-magnitude combination used in this analysis.  
We also calibrate models for the  Swope/SITe3 and MPG/ESO WFI data separately, as we have not transformed the natural Swope photometry to standard \emph{BVRI} passbands. 
The large number of clusters with spectroscopic redshifts\footnote{Spectroscopic redshifts were measured or identified from the literature for 24 clusters observed with the MPG/ESO telescope, 56 clusters imaged with the  Swope telescope,  and 100 clusters observed with the larger aperture telescopes. Some clusters with spectroscopic redshifts were observed with multiple instruments to facilitate calibration of red-sequence models.} enables these independent calibrations. 

We find a linear remapping of model redshifts, $z_\mathrm{model}$, to
spectroscopic redshifts, $z_\mathrm{spec}$,
\begin{equation}
  z_\mathrm{spec} =  A \ z_\mathrm{model} + B 
\end{equation}
is sufficient for tuning the \emph{$g - r$} vs.~\emph{i} (or \emph{z}) relation over the
redshift range $z<0.35$ as well as for tuning the Johnson color-magnitude combinations (\emph{$B-V$} vs.~\emph{R}, \emph{$V-R$} vs~\emph{R}, \emph{$R-I$} vs.~\emph{R}).

\begin{figure}
\includegraphics[width=3.5in]{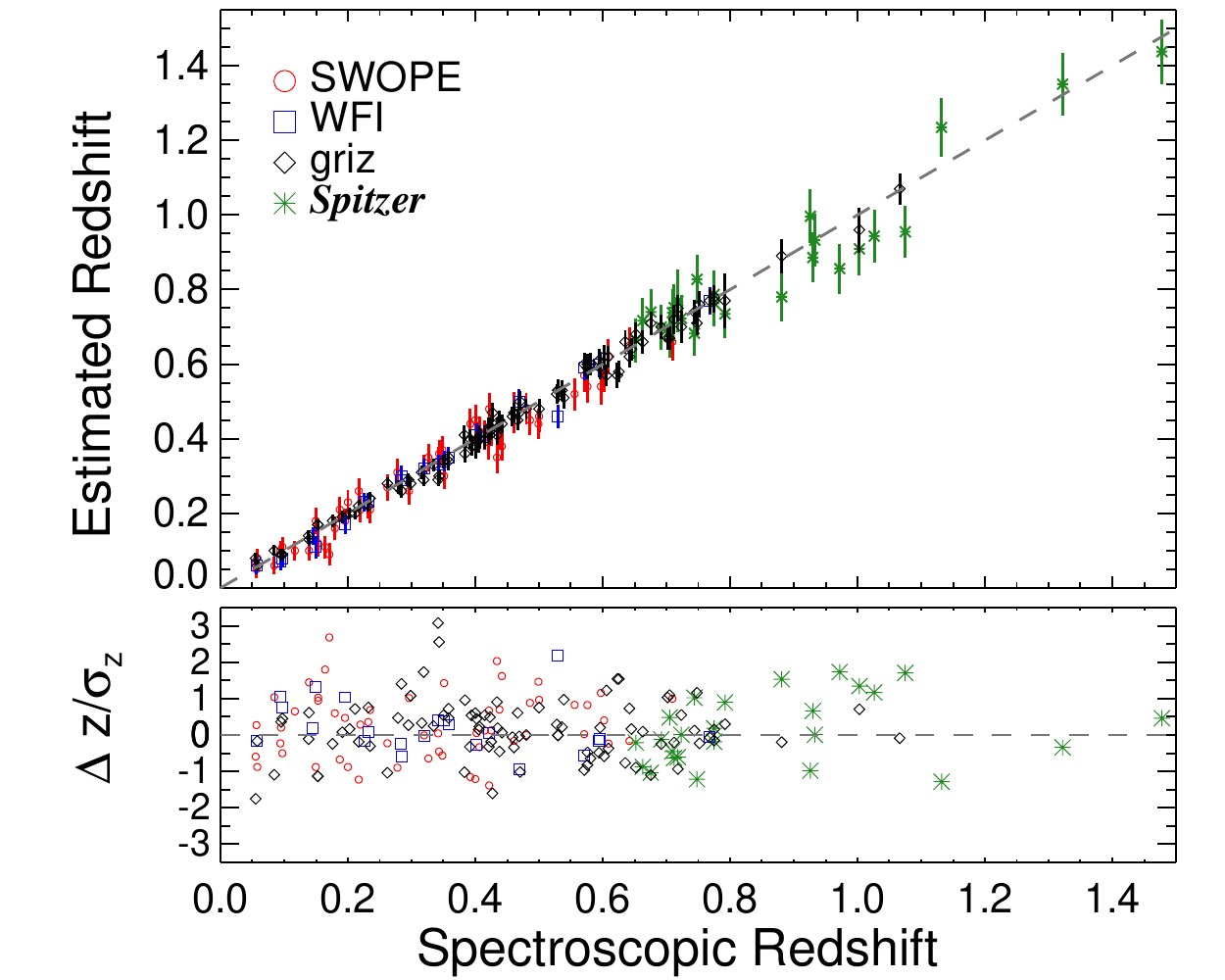}
\caption{Results of red-sequence model calibration. (Top) Photometric redshift, $z_\mathrm{est}$,  versus spectroscopic redshift, $z_\mathrm{spec}$,  for 129 spectroscopically confirmed SPT clusters. 
We plot the aggregate model tunings for  Swope/SITe3, MPG/ESO WFI, the larger class telescopes (for which the model was calibrated across instruments as all clusters were observed in the \emph{griz} system), and \spitzer/IRAC. 
Some clusters are plotted multiple times (at most once per model) as they were observed with multiple telescopes to calibrate the various redshift models. 
 (Bottom) Distribution of redshift residuals $\Delta z / \sigma_{z} = (z_\mathrm{spec} - z_\mathrm{est})/\sigma_{z_\mathrm{est}}$. The typical redshift uncertainty, $\sigma_{z}$, scales as $\sim0.013- 0.018(1+z)$ for redshifts estimated using combinations of  \emph{griz} filters, $ \sim0.021(1+z)$ for clusters imaged with the MPG/ESO WFI, $ \sim0.025(1+z)$ for Swope/SITe3,  and as $\sim\scatterspitzer (1+z)$ for redshifts determined using \spitzer /IRAC.  }
\label{fig:redshift_test}
\end{figure}

However, as noted in \citet{bleem15a}, we observe large residuals when applying such a first order correction over the broad redshift range sampled by the \emph{r,i,z} filters.    
As we expect the remapping from raw to calibrated redshifts to be smoothly varying and monotonic, we use non-linear least squares minimization to fit the $z_\mathrm{model}$ and $z_\mathrm{spec}$
relation to a monotonic function. 
This function is generated using the methodology of
\citet{ramsay98} where we have chosen sines and cosines as the basis
functions and include these functions to the 4th order. 
As in S12, we estimate the uncertainty in our model calibration by determining the quantity $\sigma_{z}$ such that the reduced chi-squared statistic,
$\chi_\mathrm{red}^{2}$ :

\begin{equation}
 \chi_\mathrm{red}^{2}  = \frac{1}{\nu}  \sum \frac{ (z_\mathrm{est} - z_\mathrm{spec})^{2}}{ (\delta_{z}(1 + z))^{2} }  = 1
\end{equation}
where $z_\mathrm{est}$ is our calibrated redshift and $\nu$ is
the number of degrees of freedom. Here the total degrees of freedom
are reduced by 2 by the linear rescaling and by 10 for higher order rescaling. 
We find  \mbox{$\delta_{z} \sim$0.025} for Swope/SITe3 , \mbox{$\delta_{z}\sim$0.021} for WFI and 
\mbox{$\delta_{z}\sim$0.013 -- 0.018} for the \emph{griz}-based redshift models. 
We plot the results of the redshift tuning in Figure \ref{fig:redshift_test}.

\subsection{Near-Infrared Redshifts}
\label{sec:nirredshift}
\begin{figure*}[hb]  
\begin{center}
\includegraphics[width=7in]{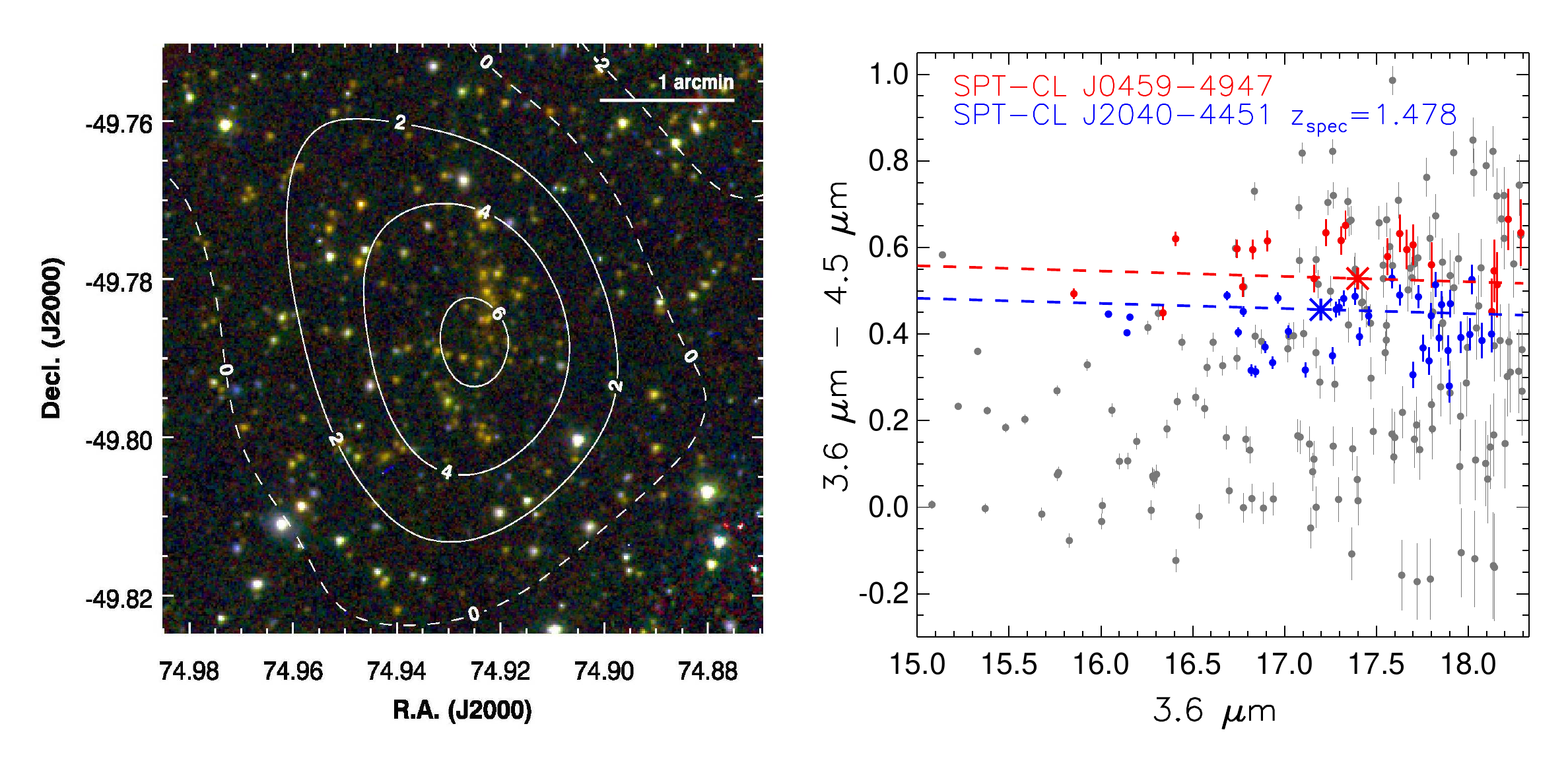}
\caption{(Left) SPT-CL~J0459$-$4947, one of three SPT clusters with an estimated redshift of $z > 1.5$ (\emph{rgb}: \spitzer/IRAC 4.5,  3.6 $\mu m$, Magellan/IMACS \emph{z}-band; over-plotted are the contours of the  SZ detection).   (Right) \spitzer \ color-magnitude diagram (with magnitudes relative to Vega): plotted in gray are all galaxies in the \spitzer \ field, over-plotted in red are the galaxies identified with the SZ detection.  
The galaxies of this massive system  (\mbox{$M_{500c}\sim 3 \times 10^{14} M_\odot h_{70}^{-1}$}) have significantly redder  \spitzer \ colors than spectroscopically confirmed SPT-CL~J2040$-$4451 at $z=1.478$ (blue), supporting that its redshift is greater than $z=1.5$. 
The color-magnitude relations of the best-fit model redshifts are over-plotted as dashed lines and the locations of model L$^{*}$ galaxies  are indicated via ``$\ast$''.
The best-fit redshift is $z=1.7 \pm 0.2$, but the model is poorly calibrated at such high redshifts.  }
\label{fig:highzcmd}
\end{center}
\end{figure*}

We analyze the NIR data as in ``Method 1'' from S12 (see \S 3.1 in that work) by supplementing the optical with the $JHK_{s}$ and IRAC imaging.  The IRAC imaging is only used in the cases where the optical imaging is not deep enough to confirm a cluster and measure the redshift.  The $JHK_{s}$ data are used as additional filters to measure at least one color across the 4000\AA\ break.  It should be noted that the \spitzer-only $[3.6]-[4.5]$ colors are probing near the peak in the stellar emission, rather than the Balmer break; as such, they are less sensitive to the effects of recent star formation or AGN activity which may be more prevalent at high redshift \citep{brodwin13}.  We have demonstrated with spectroscopic follow-up that this measurement is reliable in the redshift range relevant for confirmation of high-$z$ clusters (\citealt{stalder13,bayliss14}, and see Figure  \ref{fig:redshift_test}).  There are three clusters with $[3.6]$--$[4.5]$ galaxy colors consistent with redshifts greater than 1.5; as our models have not been tested with spectroscopic data in this redshift range, we report a lower limit of  $z=1.5$  for the redshifts of these systems (see Figure \ref{fig:highzcmd}).

\subsection{Redshift limits}
As all of our optical/NIR observations have finite depth, it is not possible to definitively rule out the existence of undiscovered, high-redshift counterparts for our unconfirmed cluster candidates. 
We instead report for each unconfirmed candidate the highest redshift for which we would have detected the overdensity of red galaxies we require to confirm a cluster. 
The depth of our follow-up imaging varies among candidates, so 
this limit is computed individually for each candidate. 
For every candidate, we determine the redshift for which a $0.4L_{*}$ red-sequence galaxy matches the measured 10$\sigma$ magnitude limit of the imaging data.
 As we require two filters to measure a redshift, we obtain the ``redshift limit'' from the second deepest of the imaging bands used in the red-sequence overdensity search. 
 A detailed description of this  procedure is provided in S12.

\begin{figure*}[t]
\begin{center}
\includegraphics[width=7in]{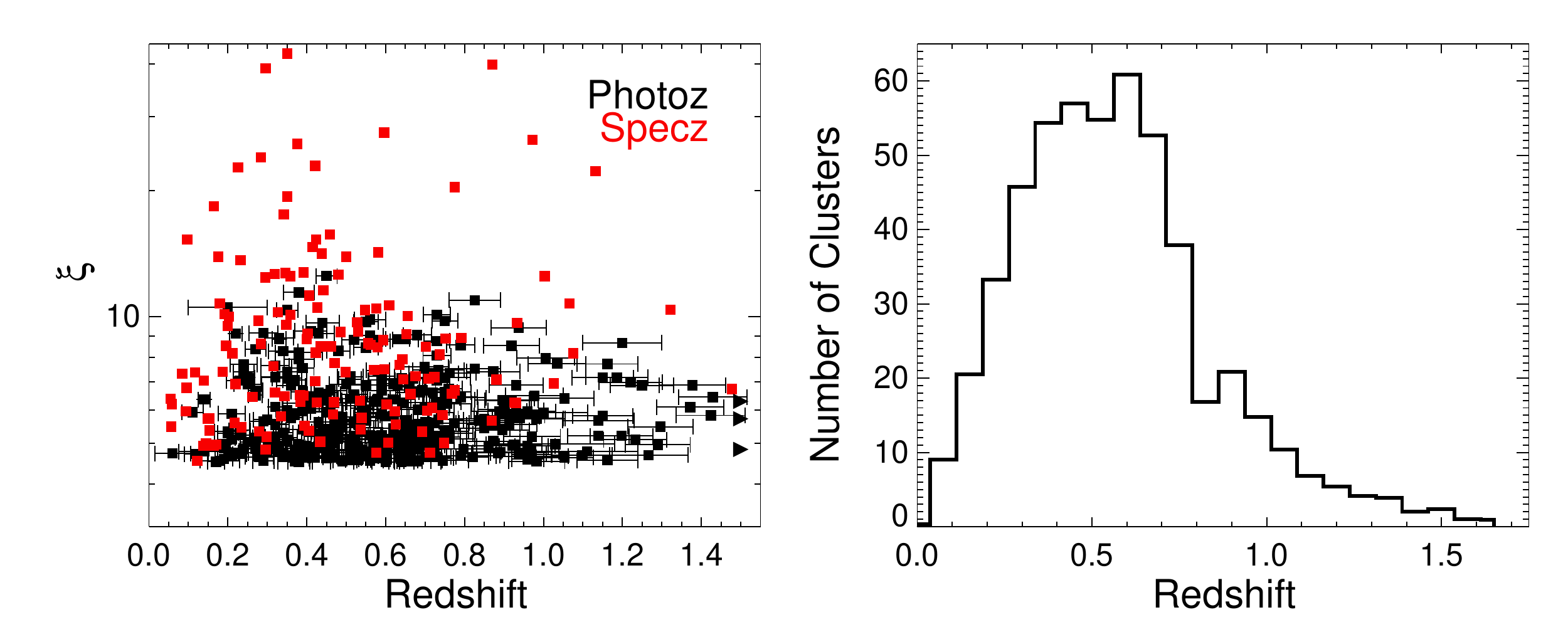}
\caption{(Left) Plot of maximum detection significance, $\xi$,   versus redshift for the confirmed SPT-SZ cluster sample; black points correspond to systems with photometrically estimated redshifts while red points represent spectroscopically confirmed clusters. We report lower limits for the redshifts of the three highest-redshift systems (see \S\ref{sec:nirredshift}). (Right) The redshift distribution of the confirmed cluster sample; the median redshift of the sample is $z=\medianz$.  The histogram does not have integer values as clusters with photometric redshift uncertainties were distributed amongst the appropriate bins.}
\label{fig:observables}
\end{center}
\end{figure*}

\section{Cluster Catalog}
\label{sec:catalog}
In Table \ref{tab:catalog} we present the complete sample of galaxy cluster candidates detected at $\xi \ge 4.5$ in the 2500 \degs \ SPT-SZ survey. 
For each candidate we provide the position, integrated Comptonization within an $0\farcm75$-radius aperture  (\S \ref{sec:ysz}), the candidate detection significance $\xi$ at the filter scale that maximizes detection significance, 
the value of the $\beta$-model core radius $\theta_c$ at which $\xi$ is reported, 
the estimated mass and  redshift (spectroscopic where available)  for confirmed clusters and redshift limit for unconfirmed candidates. We discuss the estimation of these cluster masses in \S \ref{sec:mass}.
In Figure \ref{fig:observables} we plot the  SZ detection significance versus redshift for each confirmed cluster as well as the redshift distribution of the confirmed cluster sample. 

Simulations predict that this catalog should contain only a small number of false detections
(see \S \ref{sec:simulations} for details of the simulations), and this prediction is borne out 
by our optical/NIR follow-up observations (\S \ref{sec:followup}).
Our simulations predict 18.5 false detections above $\xi=5$ for the full survey---corresponding
to a predicted purity of $95\%$ for the 402 $\xi \ge 5$ candidates in our catalog---and \nfalsefourf \ false 
detections in the full $\xi \ge 4.5$ sample---corresponding to a predicted purity of $75\%$ for the
full sample of \ncand\ cluster candidates.
Under the assumption that there are no false associations between the identified optical/NIR galaxy
overdensities and SPT detections (we estimate that  $<4\%$ of candidates will have such a false association---see discussion in \S 4.2 in S12), 
the measured cumulative purity of the sample is in excellent agreement with simulations: the purity is $\ge$\purityfive \ at $\xi \ge 5$ and  $\ge$\purityfourfive \ for the entire sample at  $\xi \ge 4.5$.
Here we quote the purity as a lower limit, as unconfirmed candidates may be clusters at redshifts too high to be confirmed with our follow-up imaging (and some lower-significance candidates have not yet been imaged).

\begin{figure*}
\begin{center}
\includegraphics[width=7in]{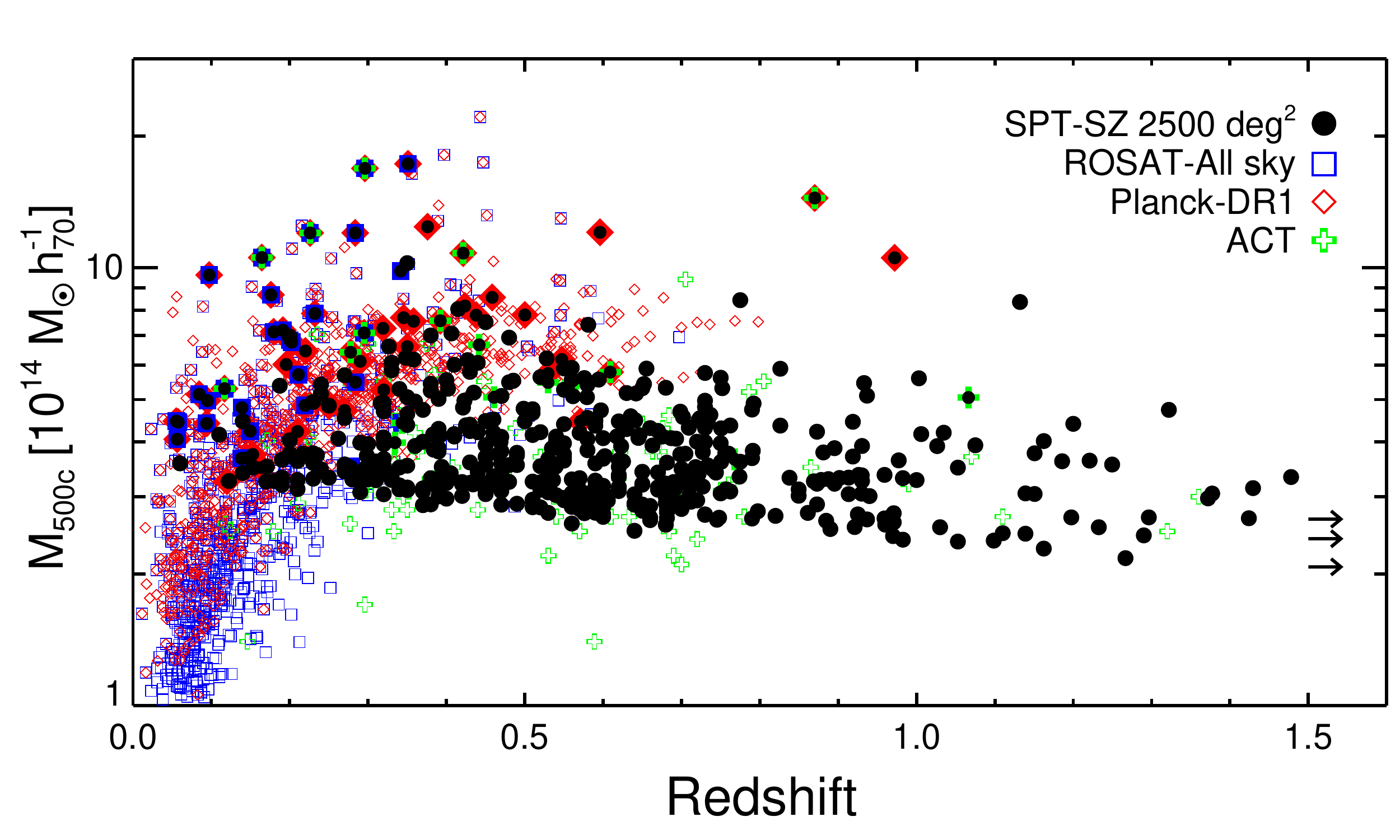}
\caption{Comparison of the 2500 \degs \ SPT-SZ cluster catalog to other X-ray and SZ-selected cluster samples. Here we plot the estimated mass versus redshift for the \nconfirm \ optically confirmed clusters from the SPT catalog, 91 clusters from the  ACT survey \citep{marriage11b,hasselfield13}, 809 SZ-selected clusters from the \planck \ survey \citep{planck13-29}, and 740 X-ray clusters selected from the \rosat \ all-sky survey \citep{piffaretti11} with $\mass \ge 1\times 10^{14}  \ h_{70}^{-1}\ \msun$. 
We mark  68\%-confidence lower limits for the redshifts of the three high-redshift SPT systems for which the \spitzer \ redshift model is poorly constrained (right arrows).  We plot clusters in common between SPT and the other datasets (see e.g., Table \ref{tab:otherid}) at the SPT mass and redshift and, for common clusters in the other datasets, at the mass and redshift of the dataset in which the cluster was first reported. 
While the SPT data provides a nearly mass-limited sample, the cluster samples selected from \rosat \ and \planck \ data are redshift-dependent owing to cosmological dimming of X-ray emission and the dilution of the SZ signal by the large \planck \ beams, respectively.}
\label{fig:comparison}
\end{center}
\end{figure*}

The SPT-SZ cluster sample contains massive galaxy clusters over a wide redshift range.
The median mass of the sample is $\mass \medianm$, and the median redshift is $z=\medianz$. 
The sample extends from $0.047 \le z  \lesssim 1.7$, and the mass threshold of the catalog is 
nearly independent of redshift (see Figure \ref{fig:comparison}).
This implies that the catalog reported here contains all of the most massive clusters in the 
$\sim$1/16th of the sky imaged by the SPT.

\begin{figure}[htb]
\begin{center}
\includegraphics[width=3.5in]{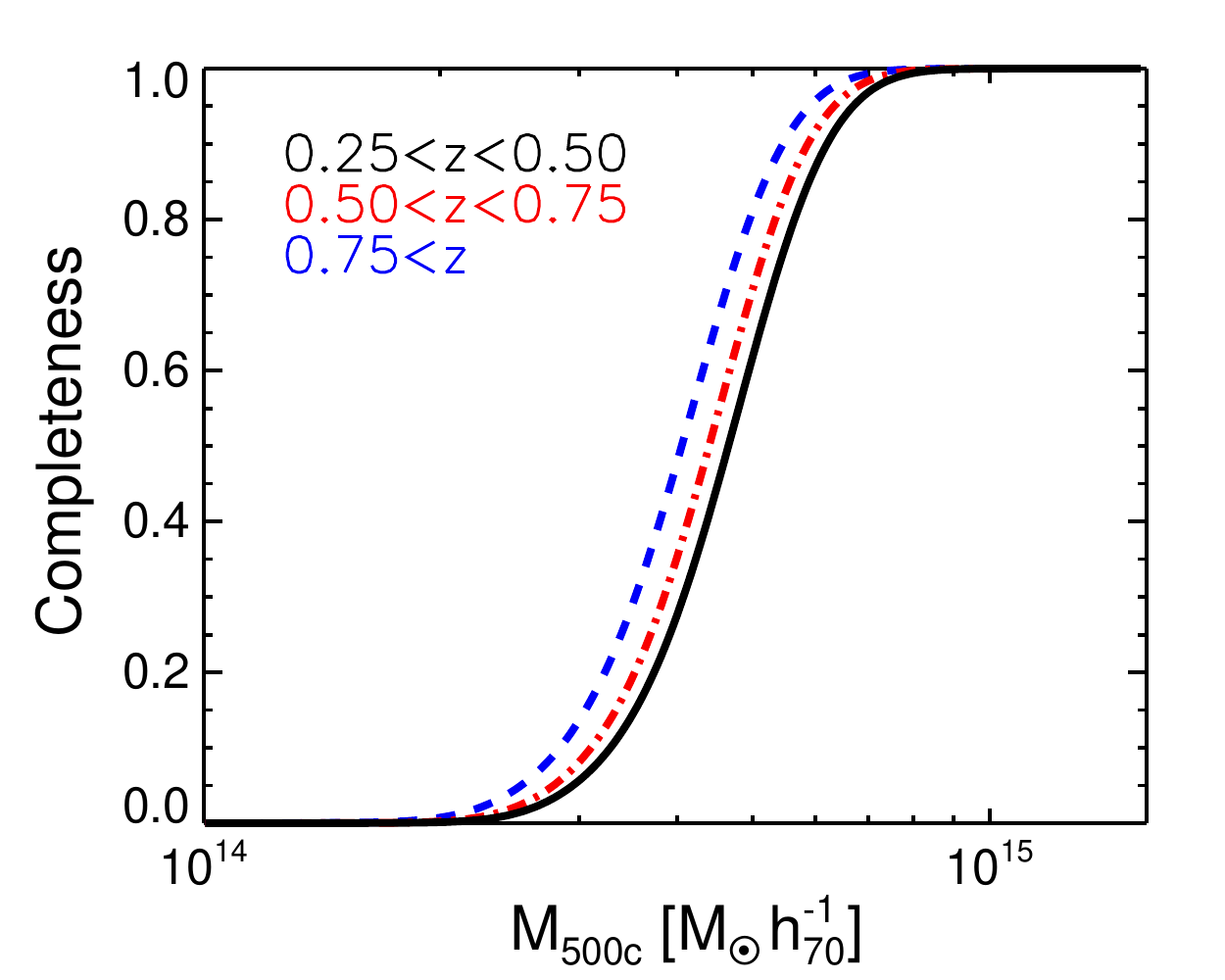}
\caption{Completeness fraction as a function of mass for the SPT cluster sample in three different redshift bins: $0.25<z<0.5$ (solid black), $0.5<z<0.75$ (dot-dashed red), $z>0.75$ (dashed blue).
The SPT sample is expected to be nearly 100\% complete for $\mass >  7 \times 10^{14} \ h_{70}^{-1}\ \msun$ at $z>0.25$. Masses are calculated for a fiducial  flat \lcdm{} cosmology with 
$\sigma_8=0.80$, $\Omega_b = 0.046$, $\Omega_c = 0.254$, $H_0 = 70$ km/s/Mpc, $\tau = 0.089$, and $n_s(0.002) = 0.972$.   Adopting the best-fit \planck  \ cosmology  \citep{planck13-16} shifts the mass thresholds up $\sim$17\%. }

\label{fig:selection}
\end{center}
\end{figure}

\begin{figure*}[htbp]
\begin{center}
 \includegraphics[width=6.25in]{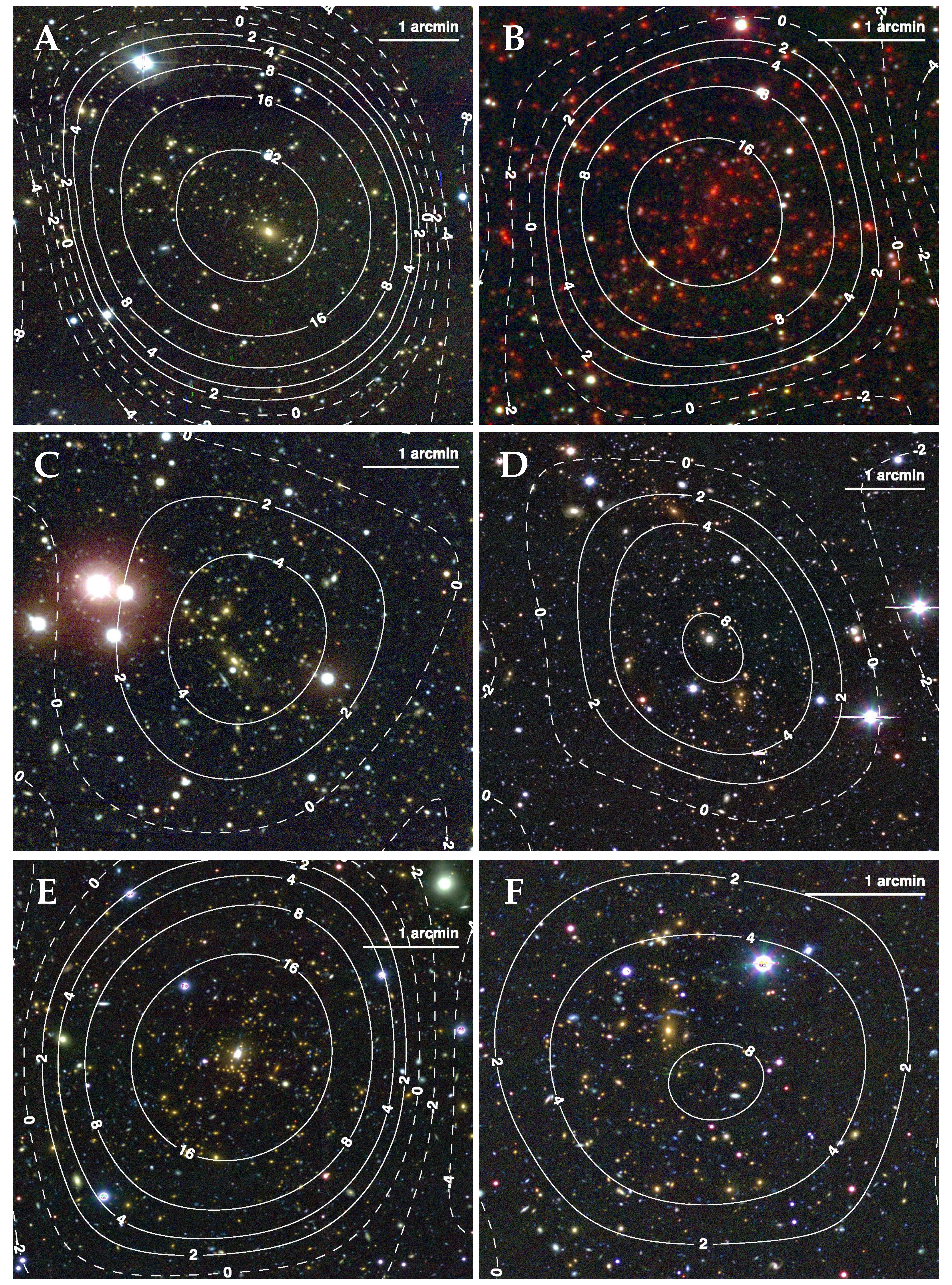}
\caption{A sample of clusters from the 2500 \degs \ SPT-SZ cluster catalog. For each cluster we display an optical/NIR \emph{rgb} image with the SZ detection contours over-plotted; see \S \ref{sec:notable} for more details on particularly notable systems. 
(a) SPT-CL~J2248$-$4431 (ACO S1063; $\xi=42.4$, $z=0.351$). This cluster is the most significant detection in the SPT sample (MPG/ESO WFI \emph{IRV}-band image). 
(b) SPT-CL~J2106$-$5844 ($\xi=22.2$, $z=1.132$)---also shown in SPT mm-wave data in Figure \ref{fig:fourpanel}---is the most massive known cluster at $z > 1$. (\spitzer/IRAC 3.6 \um, Magellan/FourStar \emph{J}-band, Magellan/IMACS \emph{i}-band image) 
(c) SPT-CL~J0410$-$6343 ($\xi=5.6$, $z=0.52$) is a  ``typical'' SPT cluster at approximately the median redshift and $\xi$ of the confirmed cluster sample. (Blanco/MOSAIC-II \emph{gri}-band image). 
(d) SPT-CL~J0307$-$6225 ($\xi=8.5$, $z=0.581$) is undergoing a major merger. As SZ selection is not greatly influenced by mergers or complicated astrophysics at the cores of clusters (e.g., \citealt{motl05}, \citealt{fabjan11}),  the SPT sample is representative of the entire population of massive clusters (Magellan/Megacam \emph{gri}-band image).
(e) SPT-CL~J2344$-$4243 (the ``Phoenix Cluster''; $\xi=27.4$, $z=0.596$) is the most X-ray luminous cluster known. We confirm this cluster as a strong lens using newly-acquired Megacam imaging (Magellan/Megacam \emph{gri}-band image).
(f) SPT-CL~J0307$-$5042 ($\xi=8.4$, $z=0.55$) is one of many strong-lensing clusters in the SPT sample (Magellan/Megacam \emph{gri}-band image). 
}
\label{fig:clusterpics}
\end{center}
\end{figure*}

In Figure \ref{fig:selection}, we show the estimated selection function of the cluster sample in three redshift bins. 
Because $\xi$ is the selection variable, the selection function can 
simply be written as the Heaviside step function $\Theta(\xi - 4.5)$. Given the 
$\xi$-$M$ relation discussed in \S\ref{sec:mass}, we transform this function to mass space, where it 
represents the probability of a cluster of a given mass to be included in the SPT-SZ sample. 
Note that at $\mass >  7 \times 10^{14} \ h_{70}^{-1} \ \msun$ and $z>0.25$, the SPT-SZ 
cluster catalog presented in this work is highly complete, meaning that nearly 
every such cluster in the surveyed area is present in the catalog. 
In Figure \ref{fig:comparison}, we compare the mass and redshift distribution of the SPT sample to those from other large cluster catalogs selected via their ICM observables: namely the clusters detected in the all-sky \rosat \ survey \citep{piffaretti11}, which includes the NORAS \citep{bohringer00}, REFLEX \citep{bohringer04}, and MACS \citep{ebeling01} cluster catalogs; the 861 confirmed clusters from the all-sky \planck \ survey  \citep{planck13-29}; and the 91 clusters that comprise the ACT cluster sample \citep{marriage11b,hasselfield13}.

The mass threshold of the SPT sample declines slightly as a function of redshift owing to a combination of effects.
At low redshifts ($z<0.3$),   increased power at large angular scales from primary CMB fluctuations 
and atmospheric noise raises  the mass threshold for a fixed $\xi$ cutoff \citep[see e.g.,][]{vanderlinde10},
while at higher redshifts the detectability of clusters is enhanced owing to increased temperatures for clusters of fixed mass.
However, both of these trends are shallow, and the nearly redshift-independent selection function of the SPT catalog stands in contrast to the
strong redshift dependence in X-ray catalogs and the \planck\ sample.
The mass threshold for X-ray catalogs is redshift-dependent owing to cosmological dimming of the X-ray emission, while the redshift dependence of the \planck \ sample is driven by the dilution of the small angular-scale signal of high-redshift clusters by the large \planck\ beam  (7\arcmin \ at 143 GHz). 

We search the literature for counterparts to SPT candidates.  We query the SIMBAD\footnote{http://simbad.u-strasbg.fr/simbad} and NED\footnote{http://nedwww.ipac.caltech.edu/} databases as well as the  union catalog of SZ sources detected by \planck \ \citep{planck13-29} for counterparts.
For confirmed clusters with $z\le 0.3$ we utilize a 5\arcmin \ association radius; otherwise we match candidates within a 2\arcmin \ radius. 
All matches are listed in Table \ref{tab:otherid}; we discuss potential false associations in the footnotes of this table. 
Additionally, we associate the brightest cluster galaxies in two clusters (SPT-CL~J0249$-$5658 and SPT-CL~J2254$-$5805) with spectroscopic galaxies from the 2dF Galaxy Redshift Survey \citep{colless03} and the 6dF Galaxy Survey \citep{jones09}, respectively. 
In total, 115 of the SPT candidates are found to have counterparts in the literature (14 of these clusters were first discovered in SPT data). 
We report the new discovery of \newcluster \ clusters here,  increasing the number of clusters first discovered in SPT data to \sptcluster. 
We highlight particularly noteworthy systems below,
and a subset of the SPT cluster catalog is shown in Figure \ref{fig:clusterpics}.  

\subsection{Cluster Mass Estimates}
\label{sec:mass}

We provide estimated masses  for all confirmed clusters in Table \ref{tab:catalog}. 
These estimates, determined from each cluster's  $\xi$ and redshift,  are based upon the methodology presented in \citet{benson13}  and R13 but are reported here for  a fixed flat \lcdm{} cosmology---with 
$\sigma_8=0.80$, $\Omega_b = 0.046$, $\Omega_m = 0.30$, $h = 0.70$, $\tau = 0.089$, and $n_s(0.002) = 0.972$---and a fixed $\xi$-mass scaling relation.
In this section we provide a brief overview of the method; readers are referred to the earlier publications for additional details.

To estimate each cluster's mass, we compute the posterior probability density function:
\begin{equation}
P(M|\xi) \propto \left. \frac{dN}{dMdz}\right|_z P(\xi|M)
\label{eq:mass_estimation}
\end{equation}
where $\frac{dN}{dMdz}$ is our assumed cluster mass function \citep{tinker08}, and $P(\xi|M)$ denotes the $\xi$-mass scaling relation. 
We assume an observable-mass scaling relation of the form
\begin{equation}
\zeta = A_\textrm{SZ} \left( \frac{\mass}{3 \times 10^{14} M_{\odot} h^{-1}} \right)^{B_\textrm{SZ}} \left(\frac{H(z)}{H(0.6)}\right)^{C_\textrm{SZ}},
\label{eqn:zetam}
\end{equation}
parameterized by the normalization $A_\textrm{SZ}$ (corrected field-by-field for the different noise levels in each field, see Table \ref{tab:fields} and R13), the slope $B_\textrm{SZ}$, the redshift evolution $C_\textrm{SZ}$ (where $H(z)$ is the Hubble parameter), and a log-normal 
scatter on $\zeta$, $D_\textrm{SZ}$, where  $\zeta$ is the ``unbiased SPT-SZ significance\footnote{See Appendix B in \citet{vanderlinde10}}"
\begin{equation}
\zeta = \sqrt{\langle\xi\rangle^2-3}
\end{equation} for $\zeta > 2$. 

We fix the scaling relation parameters to  $A_\textrm{SZ}=4.14$, $B_\textrm{SZ} = 1.44$, $C_\textrm{SZ} = 0.59$, and $D_\textrm{SZ} = 0.22$.  
These values are the best-fit weighted averages as determined from a Monte Carlo Markov chain analysis of the R13 data set assuming a fixed scatter of 0.22 and the above canonical cosmology.\footnote{This scatter was chosen to be consistent with previous constraints from X-ray measurements (\citealt{benson13}; R13).}
We caution that the masses for low-redshift clusters ($z<0.25$) may be underestimated, and for low-significance clusters ($4.5<\xi<5$) the mass estimates should be 
considered only approximate.  At low redshift the SZ signal becomes CMB-confused and therefore fails to 
obey the power-law form of the scaling relation \citep{vanderlinde10}.  
There is a more subtle complication for low-significance clusters.
When we compute $P(M|\xi)$ in Equation \ref{eq:mass_estimation}, the theoretical halo mass function is used as the Bayesian prior.  
This choice implies a one-to-one mapping between halos and $\xi$ values. 
However,  this assumption breaks down for lower-mass halos, the total number of which approaches the number of independent resolution elements in the filtered SPT maps. 
Consequently, the contribution of these lower mass systems to $P(M|\xi)$ is overestimated. 
As we have already confirmed the existence of a massive system by requiring
a significant red-sequence galaxy overdensity at the cluster location, we place a prior of $M_{500c} >  1\times 10^{14} M_\odot h^{-1}$ when computing mass estimates. 
Decreasing this prior to   $M_{500c} >  5\times 10^{13} M_\odot h^{-1}$
typically shifts the mass of the lowest significance systems by less than $0.2 \sigma$.

Because we have used a fixed cosmology and scaling relation, the uncertainty reported on each 
cluster's mass estimate only includes the contributions from measurement noise and the intrinsic
scatter in the mass-observable relation.
We also expect a comparable level of systematic 
uncertainty due to uncertainties in cosmology and scaling-relation parameters.
This systematic uncertainty will be largely correlated between clusters, and is dominated by the uncertainty in the $\xi-$mass
relation and our choice of external cosmological datasets.
Here we have intrinsically linked the cluster mass estimates to our chosen cosmology by requiring the measured R13 cluster abundance to be 
consistent with this model. 
Assuming different cosmologies can shift the cluster mass scale at a level 
comparable to the statistical uncertainty on the mass estimates.
For example, adopting the best-fit \lcdm{} model determined in R13  lowers the mass estimates by $\sim$8\% on average, whereas assuming parameter values consistent with the CMB data  from WMAP9 ($\sigma_8=0.83$, $\Omega_m=0.28$, $h=0.70$; \citealt{hinshaw13}) or \planck \ ($\sigma_8=0.84$, $\Omega_m=0.32$, $h=0.67$; \citealt{planck13-16}) typically increases the mass estimates by $\sim$4 and 17\%, respectively.

We would prefer to observationally calibrate 
the cluster scaling relations and to independently constrain cosmological parameters using clusters.  To achieve this goal, 
the SPT collaboration has undertaken a multi-wavelength campaign to obtain X-ray, galaxy velocity dispersion, and weak lensing measurements for $\sim$50-100 clusters per technique.
Early results  from this work have been presented in other SPT publications, including \citet{benson13}, R13, and \citet{bocquet15}.  
In dH14 we will present cosmological constraints from the full 2500 \degs\ SPT-SZ cluster sample. 
This analysis will combine SZ and X-ray observations  (we have obtained \chandra \  data for a significant subset of clusters, see Table \ref{tab:catalog} and \S\ref{sec:xvp})
to both constrain cosmological parameters and better quantify the systematic uncertainties in mass estimates for the cluster sample.

\subsection{Cluster Candidates in the Point-source-masked Regions}
\label{sec:clustersinmask}

The point-source veto discussed in  \S\ref{sec:extract} rejects any 
cluster detections within $8'$ of 
an emissive source detected above $5 \sigma$ at 150~GHz in SPT-SZ data. 
A total area of \areamasked\ square degrees
was excluded from cluster finding for this reason.
While such a conservative approach is appropriate when the goal is a 
cluster catalog with a clean selection function and a mass-observable relation with minimal 
outliers, it will almost certainly result in some massive clusters being excluded from
the catalog. Assuming no spatial correlation between emissive sources and clusters,
we would expect roughly 25 missed clusters above $\xi = 5$.

As in R13, we re-ran the cluster-finding algorithm only masking sources above 
$S_{150 \mathrm{GHz}}=100$~mJy (as opposed to the normal threshold
of $\sim$6~mJy).
Each detection with $\xi \ge 5$ and with no counterpart in the original, 
conservatively masked catalog was visually inspected.
The vast majority of these detections were rejected as obvious point-source-related 
artifacts, but some were clearly significant SZ decrements only
minimally affected by the nearby source.  These objects are listed in Table
\ref{tab:missed}.
We find \ncandpsmask\ objects above $\xi = 5$, within roughly $1 \sigma$ of the
naive expectation, assuming purely Poisson statistics. All six candidates that were 
identified in the analogous procedure in R13 were also identified in this 
search.\footnote{There is a systematic $\sim$$0\farcm75$ offset between the 
positions of the R13 clusters found in source-masked areas and their counterparts
in Table \ref{tab:missed}. This is due to a small error in the position calculation in R13.
That error was only present in the source-masked cluster positions; as noted in 
\S \ref{sec:compare}, the positions in the main catalog here and the main catalog in R13 are in excellent agreement.}

As in R13, these auxiliary candidates were not included in our optical/IR follow-up 
campaign, and
they are not included in the cosmological analysis or in the total number of 
candidates quoted in the rest of the text. We have searched the literature
for counterparts to these clusters in other catalogs,
and any literature counterparts are listed in Table \ref{tab:missed}. We 
also searched the RASS bright- and faint-source catalogs \citep{voges99,voges00}  for X-ray counterparts, 
and we show in Table \ref{tab:missed} whether we find a RASS counterpart within either
a 2\arcmin \ or 5\arcmin \ radius.

\subsection{Comparison to Previous SPT Catalogs}
\label{sec:compare}
The SPT collaboration has published three previous samples of galaxy clusters (\citealt{vanderlinde10},
\citealt{williamson11}, R13). The catalog we present here encompasses all of the data used in these 
previous analyses, so it is a potentially useful cross-check of our new analysis to compare to these earlier samples. 
The R13 sample included all the clusters in \citet{vanderlinde10}, 
using the exact same data and analysis, so we do not perform a separate comparison to \citet{vanderlinde10}
here. 
The catalog published in R13 was constructed using an analysis nearly identical to that used 
here;  some small differences were pointed out in \S \ref{sec:mapmaking} and we describe the remaining differences here. 

First, there is a small difference in the area in each field's map over which clusters are extracted.
This area is generally defined in SPT analyses as the set of pixels with total weight above a given fraction of the median
weight in the map. 
To ensure full coverage of the 2500 \degs\ region (i.e., no gaps between fields), 
we use a slightly lower threshold in this work ($70\%$ of the median weight, as opposed to $80\%$ in R13).
This results in very slight differences in noise properties and, hence, $\xi$ values for extracted 
cluster candidates. 

For the three fields observed in 2009 (\textsc{ra21hdec$-$60}, \textsc{ra3h30dec$-$60}, and
\textsc{ra21hdec$-$50}), we use the same data as in R13, with the only significant analysis difference 
being the different field border definitions. 
Therefore, we expect the list of cluster candidates in those fields to be very similar to the corresponding
list in R13.
 Indeed, above a signal-to-noise threshold of $\xi=5.5$ in these fields, the cluster lists are identical 
between this work and R13, and the ratio of $\xi$ values for these clusters between the two 
analyses is $1.00 \pm 0.04.$ 
Two clusters between $\xi=5$ and $\xi=5.5$ from R13 (SPT-CL~J0411$-$5751 and SPT-CL~J2104$-$5224)
are not included in the catalog presented here. 
This absence is due to an update in the point-source lists between the two analyses; both of the $\xi>5$ R13 clusters missing from the
current catalog are within $4^\prime$ of newly identified point sources. 
There are two clusters in these fields at $5 \le \xi \le 5.5$ in the current list that are not in the R13 sample (SPT-CL~J2158$-$5451 and 
SPT-CL~J2143$-$5509); these are a result of the redefined field boundaries.

For the two fields originally observed in 2008 and re-observed in 2010 or 2011
(\textsc{ra5h30dec$-$55} and \textsc{ra23h30dec$-$55}), the analysis in R13 
used only the 2008 data, whereas we use both years' data here. 
We thus expect the cluster list from R13 in these fields to be a subset of the list from these fields in this work, and we expect the average
$\xi$ for those clusters to be higher in this work, by a factor related to the extra 150~GHz depth and 
added 95~GHz information. 
As expected, all clusters above $\xi=6$ from R13 in these fields are 
also present in the sample from this work, and the mean $\xi$ ratio between this work and R13 
is $1.28 \pm 0.10$ for these objects. 
We note that this ratio should agree with the ``field scaling factors" used in dH14 to
rescale the normalization parameter in the $\xi$-mass relation.
The average of the two fields' scaling factors in dH14 is 
1.35; the $\xi$ ratio we determine here is 
$\sim$5\% lower than this value but consistent within $1 \sigma$.

There is one cluster above $\xi=5$ from the \textsc{ra5h30dec$-$55} and \textsc{ra23h30dec$-$55}
fields that drops out of the R13 catalog when we add the new data. The cluster candidate
SPT-CL~J2343$-$5521 (detected at $\xi=5.74$ in R13) is detected at $\xi < 4$ in the full 
two-year, two-band data. This trend of 
significance with added data is strong evidence that this is a false detection in the single-year, 
single-frequency data. This was already the preliminary conclusion in \citet{vanderlinde10}
and R13, based on null results from optical and X-ray follow-up observations of this candidate.

We also compare the best-fit positions for the clusters detected in common between R13 and
this work. The mean difference in positions of clusters in the two catalogs is 
$0\farcs15 \pm 0\farcs59$ in right ascension and $-0\farcs04 \pm 0\farcs69$
in declination.

We  also cross-check the cluster catalog presented in \citet{williamson11} with the catalog 
presented here. \citet{williamson11} used full-depth observations of 60\% of the SPT-SZ survey
area and ``preview-depth'' (roughly three times the nominal noise level) observations of the 
rest of the survey to compile a catalog of the most massive clusters over the full 2500 \degs\ 
region. All clusters detected in that work are also detected here. One cluster from 
\citet{williamson11}---SPT-CL~J0245$-$5302, also known as ACO~S0295---is near a bright radio 
source and is thus not included in the official catalog in Table \ref{tab:catalog}; it is, however, detected
in the alternative analysis in \S \ref{sec:clustersinmask} that avoids only the very brightest point sources,
and it is included in Table \ref{tab:missed}. The values of $\xi$ in this work for the clusters from
full-depth data in \citet{williamson11} are consistent with the $\xi$ values reported in that work; 
for clusters found in preview-depth data in \citet{williamson11}, the $\xi$ values reported here
have increased by roughly a factor of two over that work.

Finally, we note that there may  be small differences in redshifts/redshift limits and the confirmation status of cluster candidates. 
These changes are driven by a mixture (in varying degree per candidate) of additional follow-up data, improved optical data processing and re-estimated redshifts (\S \ref{sec:redshifts}) calibrated using the enlarged SPT spectroscopic sample (\S \ref{sec:specsample}).

\begin{center}
\begin{deluxetable*}{ lrrccccc}
\tablecaption{\label{tab:missed}Cluster candidates above $\xi=5$ in the source-masked area}
\tablehead{
\multicolumn{1}{c}{} &
\multicolumn{4}{l}{} &
\multicolumn{2}{c}{X-ray counterpart?} &
\multicolumn{1}{c}{}  \\
\colhead{SPT ID} & 
\colhead{R.A.} & 
\colhead{DEC}  & 
\colhead{\nSN{}} & 
\colhead{$\theta_\textrm{c}$} & 
\colhead{within $2^\prime$} &
\colhead{within $5^\prime$} &
\colhead{Literature name} \\
}
\startdata
\input{tab_missed_2500d.tex}
\enddata
\tablecomments{ Cluster candidates identified in a version of the
analysis in which only the very brightest ($> 100$~mJy) point sources are masked
(see text for details).
Only candidates from the area masked in the standard analysis are listed here.  
These candidates are not included in cosmological analyses or in the candidate numbers
 quoted in the text. X-ray counterparts are searched for in the RASS
bright- and faint-source catalogs. Literature name and reference are given for the
first known identification of the cluster.
}
\tablenotetext{a}{\citet{abell89}}
\tablenotetext{b}{\citet{marriage11b}}
\end{deluxetable*}
\end{center}

\subsection{Notable Clusters}
\label{sec:notable}
In this section we highlight particularly notable clusters and subsets of clusters from the SPT-SZ cluster catalog. 

\subsubsection{The SPT-XVP Sample} 
\label{sec:xvp}
Eighty of the most significant SPT clusters discovered in the first 2000 \degs \  at  $z>0.4$ have been observed by \chandra \ as part of a large X-ray Visionary Project (XVP; PI Benson). 
As described in R13 and dH14, these observations, with $\sim$2000 X-ray counts/cluster, play a critical role in constraining the SZ-mass scaling relation and greatly strengthen the cosmological constraining power of the  SPT cluster sample. 
The broad redshift range of this dataset has also enabled constraints on the redshift evolution of the X-ray properties of massive clusters, including measurements of their cooling properties \citep{mcdonald13} as well as  the redshift evolution of the temperature, pressure and entropy profiles of clusters \citep{mcdonald14}.

\subsubsection{Massive Clusters at z $>$ 1}
The nearly redshift-independent selection of the SPT-SZ cluster sample has led to the discovery of a number of massive, high-redshift clusters. 
Thirty-seven clusters reported in this work have redshifts estimated at $z>1$; three systems  reported here for the first time: SPT-CL~J0459$-$4947 ($\xi =  6.29$), SPT-CL~J0446$-$4606  ($\xi = 5.71$) and SPT-CL~J0334$-$4645 ($\xi = 4.83$) have redshifts estimated from \spitzer \ observations at $z >1.5$. 
Several of the $z > 1$ systems have been the focus of more detailed study including: 
SPT-CL~J0546$-$5345, the first $z>1$ cluster detected by its SZ signature \citep{brodwin10}; SPT-CL~J2106$-$5844, the most massive known cluster at $z > 1$ ($\mass=8.3 \times 10^{14}  \ h_{70}^{-1} \msun$; \citealt{foley11}); 
SPT-CL~J0205$-$5829 at $z=1.32$ \citep{stalder13}, which features a red sequence whose bright galaxies are well-evolved by $z=1.3$; 
and SPT-CL~J2040$-$4451, which intriguingly shows signs of active star formation \citep{bayliss14}. 
At $z=1.478$, SPT-CL~J2040$-$4451 is the highest-redshift spectroscopically confirmed SPT cluster to date. 

\subsubsection{Strong Lensing Clusters}
\label{sec:stronglens}
A number of SPT clusters can be identified from the literature and existing SPT follow-up observations as strong gravitational lenses.  
Previous SPT publications first identified SPT-CL~J0509$-$5342, SPT-CL~J0546$-$5345 \citep{staniszewski09}, SPT-CL~J0540$-$5744, SPT-CL~J2331$-$5051 \citep{high10}, SPT-CL~J2011$-$5228, and SPT-CL~J2011$-$5725 \citep{Song12b} as strong-lensing clusters using optical imaging data. 
SPT-CL~J2332$-$5358 was identified as a lens by the presence of a multiply imaged, high-redshift ($z=2.73$), dusty star-forming galaxy \citep{greve12,aravena13}.
The ACT team first reported the discovery of several clusters in the SPT sample (see Table \ref{tab:otherid}) and identified 3 of these systems as strong lenses:  SPT-CL~J0304$-$4921, SPT-CL~J0330$-$5228 \citep{menanteau10b}, and SPT-CL~J0102$-$4915 \citep{menanteau12,zitrin13}.  Other previously identified strong-lensing systems include  SPT-CL~J0658$-$5556 (1E0657-56/Bullet Cluster; \citealt{mehlert01}), SPT-CL~2031$-$4037 (RXC~J2031.8$-$4037; \citealt{christensen12}),
SPT-CL~J2248$-$4431 (ACO~1063S; \citealt{gomez12}), and SPT-CL~J2351$-$5452 (SCSO~J235055$-$530124; \citealt{menanteau10,buckleygeer11}). 

In Table \ref{tab:catalog}, we report 34 additional strong gravitational lenses.   
Optical images of two newly identified strong lenses (SPT-CL~J2344$-$4243 and SPT-CL~J2138$-$6008) are shown in Figure \ref{fig:clusterpics}. 
While the majority of these lenses have been identified in ground-based imaging from the follow-up program described in \S\ref{sec:followup}, a subset has been identified via the presence of bright arcs in deeper, higher-quality imaging acquired as part of our weak-lensing mass calibration efforts (including 16 lensing clusters in data from the {\sl{Hubble Space Telescope}}).
We note that, given the heterogeneity in image quality in existing follow-up data, this list of strong lenses represents neither an exhaustive nor a uniformly selected sample of systems. 

\subsubsection{Notable Individual Systems}
\begin{itemize}

\item SPT-CL~J2248$-$4431:  First reported as ACO~S1063, it is the most significant detection ($\xi=42.4$, $z=0.351$, $\mass=17.3 \times 10^{14} \  h_{70}^{-1} \ \msun$) in the SPT-SZ sample.  The cluster is the most massive cluster in the SPT-SZ sample, and is the second most X-ray luminous cluster in the REFLEX X-ray catalog \citep{bohringer04}.  It is scheduled for ultra-deep {\sl Hubble Space Telescope} observations, as one of its Frontier Fields.\footnote{http://www.stsci.edu/hst/campaigns/frontier-fields/}

\item SPT-CL~J0102$-$4915: First reported in \citet{menanteau10b}, this cluster is also known as ``El Gordo'' \citep{menanteau12}.
Detected in the SPT-SZ survey at $\xi=39.9$, this massive ($\mass=14.4 \times 10^{14} \  h_{70}^{-1} \ \msun$) merging cluster at $z=0.870$ is the second most significant detection in the SPT-SZ sample. It has a very high X-ray temperature (14.5 keV), and an X-ray luminosity that makes it the second most X-ray luminous cluster in the SPT-SZ sample \citep{menanteau12}. 

\item SPT-CL~J0658$-$5556: This cluster is the well-known ``Bullet'' cluster (1ES~0657$-$558; $z=0.296$, \citet{tucker98,clowe06}).  Detected at $\xi = 39.0$, this cluster is the second most massive system ($\mass=16.9 \times 10^{14} \ h_{70}^{-1} \ \msun$) in the SPT-SZ cluster sample. 

\item SPT-CL~J2344$-$4243: This system, first reported in \citet{williamson11}, is also known as the ``Phoenix Cluster'' ($\xi=27.4$,  $z=0.596$, $\mass=12.0 \times 10^{14} \ h_{70}^{-1} \ \msun$). It is the most X-ray luminous cluster known in the universe. 
The properties of this system, including those of its central galaxy which exhibits an exceptionally high rate of star formation,  are explored in detail in  \citet{mcdonald12,mcdonald13b,mcdonald14a}. We use newly acquired Magellan/Megacam imaging to identify this system as a strong lens (see Figure \ref{fig:clusterpics}). 

\item SPT-CL~J0615$-$5746: This cluster ($\xi=26.4$, $z=0.972$, $\mass=10.5 \times10^{14} \ h_{70}^{-1} \ \msun$) was first reported by SPT in \citet{williamson11}, and also appears in the {\sl Planck} cluster catalog \citep{planck11-5.1a}.  It has a measured X-ray luminosity equal to the Bullet cluster \citep{planck11-26}, but is at significantly higher redshift.  

\item SPT-CL~J2106$-$5844: As mentioned above, this cluster ($\xi=22.2$, $z=1.132$, $\mass=8.3~\times~10^{14}  \ h_{70}^{-1} \msun$) is the most massive known cluster at $z>1$.  It has a measured X-ray luminosity nearly equal to SPT-CL~J0615$-$5746, and is described in more detail in \citet{foley11}.

\end{itemize}


\section{Conclusions}
\label{sec:conclusions}
This work has documented the construction and properties of a catalog of galaxy
cluster candidates, selected via their SZ signature in the 2500 \degs\ SPT-SZ survey.
Using a spatial-spectral matched filter and a simple peak-finding algorithm, we have 
used the 95 and 150~GHz survey data to identify \ncand\  cluster candidates above a
signal-to-noise threshold of $\xi=4.5$. From simulated data, we have estimated the purity 
of this sample to be $75\%$; 
above a threshold of $\xi=5$, simulations
have indicated that the sample should be $95\%$ pure. 
In our optical/NIR follow-up data, we identified clear overdensities of similarly colored galaxies 
in the direction of \nconfirm \ (\purityfourfive) of the $\xi \ge 4.5$ cluster candidates and 
  \nconfirmfive \ (\pctconfirmfive$\%$) of the $\xi \ge 5$ cluster candidates, confirming 
the predictions from simulations.
Of these confirmed clusters, \sptcluster \ were first
identified in SPT data, including 250 new discoveries reported in this work.

We have also used the optical/NIR data to estimate photometric redshifts for all of our
candidates with clear counterparts, and we have estimated lower redshift limits for the 
candidates without counterparts. We have combined these measurements with spectroscopic 
redshifts for \nspec \ clusters in the sample to estimate the redshift distribution of the sample. 
The median redshift is $z_\mathrm{med} = \medianz$, 83 ($16\%$) of the confirmed clusters lie
at $z \ge 0.8$, and 37 ($7 \%$)  lie at $z \ge 1$. 
Using the framework developed for R13, we report masses using the best-fit $\xi$-mass relation for a fixed flat \lcdm{} \  cosmology with $\Omega_{\textrm{m}} = 0.3$, $h=0.7$ and $\sigma_{8} = 0.8$. 
The typical mass of clusters in the sample is $\mass  \medianm$, nearly 
independent of redshift.  Work is ongoing to improve the mass calibration of SPT clusters using X-ray, galaxy velocity dispersion, and optical weak lensing measurements. 
Selected data reported in this work, as well as future cluster masses estimated using these datasets, will be hosted at \webaddress. 

SZ-selected samples of galaxy clusters from data with sufficient angular resolution are expected to have a nearly redshift-independent mass limit, 
and the distribution in mass and redshift of the sample presented here is fully consistent with this expectation. 
This combination of clean selection, large redshift extent, and high typical mass make this sample of particular interest for cosmological and cluster physics analyses.

The catalog presented in this work represents the complete sample of clusters detected at high significance in 
the 2500 \degs\ SPT-SZ survey. 
The program of galaxy cluster science with the SPT
continues with the currently fielded SPTpol receiver \citep{austermann12} and will 
expand further with the expected deployment of the SPT-3G receiver \citep{benson14}.

\section*{Acknowledgements} 

The South Pole Telescope is supported by the National Science Foundation through grant PLR-1248097.  
Partial support is also provided by the NSF Physics Frontier Center grant PHY-1125897 to the Kavli Institute of Cosmological Physics at the University of Chicago, the Kavli Foundation and the Gordon and Betty Moore Foundation grant GBMF 947.
Galaxy cluster research at Harvard is supported by NSF grant AST-1009012 and at SAO in part by NSF grants AST-1009649 and MRI-0723073. The McGill group acknowledges funding from the National Sciences and Engineering Research Council of Canada, Canada Research Chairs program, and the Canadian Institute for Advanced Research. Argonne National Laboratory's work was supported under U.S. Department of Energy contract DE-AC02-06CH11357. 
This work was partially completed at Fermilab, operated by Fermi Research Alliance, LLC under Contract No. De-AC02-07CH11359 with the United States Department of Energy.  The Munich group acknowledges the support by the DFG Cluster of Excellence ``Origin and Structure of the Universe'' and the Transregio program TR33 ``The Dark Universe''. 
MM acknowledges support by NASA through a Hubble Fellowship grant HST- HF51308.01-A awarded by the Space Telescope Science Institute.
T.S. and D.A. acknowledge support from the German Federal Ministry of Economics and Technology (BMWi) provided through DLR under project 50 OR 1210.

Optical imaging data from the Blanco 4 m at Cerro Tololo Interamerican Observatories (programs 2005B- 0043, 2009B-0400, 2010A-0441, 2010B-0598) and spectroscopic observations from VLT programs 086.A-0741, 087.A-0843, 088.A-0796(A), 088.A- 0889(A,B,C), and 286.A-5021 and Gemini programs GS-2009B-Q-16,  GS-2011A-C-3, GS-2011B-C-6, GS-2012A-Q-4, GS-2012A-Q-37, GS-2012B-Q-29, GS-2012B-Q-59, GS-2013A-Q-5, GS-2013A-Q-45, GS-2013B-Q-25 and GS-2013B-Q-72 were included in this work. Additional data were obtained with the 6.5 m Magellan Telescopes and the Swope telescope, which are located at the Las Campanas Observatory in Chile  and the MPG/ESO 2.2 m and ESO NTT located at La Silla Facility in Chile.  This work is based in part on observations made with the Spitzer Space Telescope (PIDs 60099, 70053, 80012 and 10101), which is operated by the Jet Propulsion Laboratory, California Institute of Technology under a contract with NASA. Support for this work was provided by NASA through an award issued by JPL/Caltech. This work is also partly based on observations made with the NASA/ESA Hubble Space Telescope, obtained at the Space Telescope Science Institute, which is operated by the Association of Universities for Research in Astronomy, Inc., under NASA contract NAS 5-26555; these observations are associated with programs 12246, 12477, and 13412.
 The Digitized Sky Surveys were produced at the Space Telescope Science Institute under U.S. Government grant NAG W-2166. The images of these surveys are based on photographic data obtained using the Oschin Schmidt Telescope on Palomar Mountain and the UK Schmidt Telescope. The plates were processed into the present compressed digital form with the permission of these institutions.

{\it Facilities:}
\facility{Blanco (MOSAIC, NEWFIRM)},
\facility{MPG/ESO 2.2m (WFI)}
\facility{Gemini-S (GMOS)},
\facility{HST (ACS)},
\facility{Magellan:Baade (IMACS, FourStar)},
\facility{Magellan:Clay (LDSS3,Megacam)},
\facility{NTT (EFOSC)},
\facility{Spitzer (IRAC)},
\facility{South Pole Telescope},
\facility{Swope (SITe3)},
\facility{VLT:Antu (FORS2)}

\bibliographystyle{fapj}
\bibliography{../../BIBTEX/spt}

\clearpage

\appendix
\section{The Cluster Catalog}

\LongTables 
\begin{center}
\def\arraystretch{1.2}
\tabletypesize{\scriptsize}

\begin{deluxetable}{ lcc | cc |c|c|c|c|c}
\tablecaption{Galaxy clusters above $\xi = 4.5$ in $2500$ square degrees observed by the SPT.
\tablewidth{0pt}
\label{tab:catalog}}
\tablehead{
\multicolumn{1}{c}{\bf SPT ID} & 
\multicolumn{1}{c}{{\bf R.A.}} & 
\multicolumn{1}{c}{{\bf Decl.}} & 
\multicolumn{2}{c}{\bf Best} &
\multicolumn{1}{c}{{\bf Y$_{\rm SZ}$} $\times 10^6$} &
\multicolumn{1}{c}{\bf Redshift} &
\multicolumn{1}{c}{\bf M$_{500c}$}  &
\multicolumn{1}{c}{\bf Imaging}  &
\multicolumn{1}{c}{\bf Notes} \\
\colhead{} & 
\colhead{(J2000)} & 
\colhead{(J2000)}  & 
\colhead{\nSN{}} & 
\colhead{$\theta_\textrm{c}$} & 
\colhead{(${\rm arcmin}^2$)}  & 
\colhead{} &
\colhead{($10^{14} h_{70}^{-1} M_\odot$)} &
\colhead{}  &
\colhead{}
}

\startdata 
\input{final_cluster_table}
\enddata
\end{deluxetable}

\end{center}
\clearpage

\pagestyle{plain}

\pagestyle{empty}
\begin{deluxetable}{ lcccc}
\centering
\tablecaption{\label{tab:otherid}Clusters with matches in other catalogs} \small
\tablewidth{0pt}
\tablehead{
\colhead{SPT ID} & 
\colhead{First ID, ref.} & 
\colhead{All catalogs with match} & 
\colhead{$z$} & 
\colhead{Lit.~$z$, ref.} 
}
\startdata
\input{tab_other}
\enddata
\end{deluxetable}
\clearpage

\pagestyle{plain}

\end{document}

%% file: spt_authors.tex
\altaffiltext{\ANL}{Argonne National Laboratory, High-Energy Physics Division, 9700 S. Cass Avenue, Argonne, IL, USA 60439}
\altaffiltext{\KICPChicago}{Kavli Institute for Cosmological Physics,
University of Chicago,
5640 South Ellis Avenue, Chicago, IL 60637}
\altaffiltext{\PhysicsUChicago}{Department of Physics,
University of Chicago,
5640 South Ellis Avenue, Chicago, IL 60637}
\altaffiltext{\CfA}{Harvard-Smithsonian Center for Astrophysics,
60 Garden Street, Cambridge, MA 02138}
\altaffiltext{\McGill}{Department of Physics,
McGill University,
3600 Rue University, Montreal, Quebec H3A 2T8, Canada}
\altaffiltext{\UChicago}{University of Chicago,
5640 South Ellis Avenue, Chicago, IL 60637}
\altaffiltext{\KIPAC}{Kavli Institute for Particle Astrophysics and Cosmology, Stanford University, 452 Lomita Mall, Stanford, CA 94305}
\altaffiltext{\Stanford}{Department of Physics, Stanford University, 382 Via Pueblo Mall, Stanford, CA 94305}
\altaffiltext{\SLAC}{SLAC National Accelerator Laboratory, 2575 Sand Hill Road, Menlo Park, CA 94025}
\altaffiltext{\AIfA}{Argelander-Institut f{\"u}r Astronomie, Auf dem H{\"u}gel 71, D-53121 Bonn, Germany}
\altaffiltext{\MIT}{Kavli Institute for Astrophysics and Space
Research, Massachusetts Institute of Technology, 77 Massachusetts Avenue,
Cambridge, MA 02139}
\altaffiltext{\Harvard}{Department of Physics, Harvard University, 17 Oxford Street, Cambridge, MA 02138}
\altaffiltext{\FNAL}{Fermi National Accelerator Laboratory, Batavia, IL 60510-0500}
\altaffiltext{\AAUChicago}{Department of Astronomy and Astrophysics,
University of Chicago,
5640 South Ellis Avenue, Chicago, IL 60637}
\altaffiltext{\Munich}{Department of Physics,
Ludwig-Maximilians-Universit\"{a}t,
Scheinerstr.\ 1, 81679 M\"{u}nchen, Germany}
\altaffiltext{\ExcellenceCluster}{Excellence Cluster Universe,
Boltzmannstr.\ 2, 85748 Garching, Germany}
\altaffiltext{\Miss}{Department of Physics and Astronomy, University of Missouri, 5110 Rockhill Road, Kansas City, MO 64110}
\altaffiltext{\EFIChicago}{Enrico Fermi Institute, University of Chicago, 5640 South Ellis Avenue, Chicago, IL 60637}
\altaffiltext{\NIST}{NIST Quantum Devices Group, 325 Broadway Mailcode 817.03, Boulder, CO, USA 80305}
\altaffiltext{\PUC}{Departamento de Astronomia y Astrosifica, Pontificia Universidad Catolica,
Chile}
\altaffiltext{\Caltech}{California Institute of Technology, 1200 E. California Blvd., Pasadena, CA 91125}
\altaffiltext{\CIFAR}{Canadian Institute for Advanced Research, CIFAR Program in Cosmology and Gravity, Toronto, ON, M5G 1Z8, Canada}
\altaffiltext{\illast}{
Astronomy Department,
University of Illinois at Urbana-Champaign,
1002 W.\ Green Street,
Urbana, IL 61801 USA}
\altaffiltext{\illphy}{
Department of Physics,
University of Illinois Urbana-Champaign,
1110 W.\ Green Street,
Urbana, IL 61801 USA}
\altaffiltext{\Berkeley}{Department of Physics,
University of California, Berkeley, CA 94720}
\altaffiltext{\MPE}{Max-Planck-Institut f\"{u}r extraterrestrische Physik,
Giessenbachstr.\ 85748 Garching, Germany}
\altaffiltext{\UFlorida}{Department of Astronomy, University of Florida, Gainesville, FL 32611}
\altaffiltext{\Colorado}{Department of Astrophysical and Planetary Sciences and Department of Physics,
University of Colorado,
Boulder, CO 80309}
\altaffiltext{\LeidenObservatory}{Leiden Observatory, Leiden University, Niels Bohrweg 2, 2333 CA, Leiden, the Netherlands}
\altaffiltext{\Davis}{Department of Physics, 
University of California, One Shields Avenue, Davis, CA 95616}
\altaffiltext{\LBNL}{Physics Division,
Lawrence Berkeley National Laboratory,
Berkeley, CA 94720}
\altaffiltext{\Arizona}{Steward Observatory, University of Arizona, 933 North Cherry Avenue, Tucson, AZ 85721}
\altaffiltext{\Michigan}{Department of Physics, University of Michigan, 450 Church Street, Ann  
Arbor, MI, 48109}
\altaffiltext{\Minnesota}{Physics Department, University of Minnesota, 116 Church Street S.E., Minneapolis, MN 55455}
\altaffiltext{\Melbourne}{School of Physics, University of Melbourne, Parkville, VIC 3010, Australia}
\altaffiltext{\STScI}{Space Telescope Science Institute, 3700 San Martin
Dr., Baltimore, MD 21218}
\altaffiltext{\CaseWestern}{Physics Department, Center for Education and Research in Cosmology and Astrophysics, Case Western Reserve University, Cleveland, OH 44106}
\altaffiltext{\SAIC}{Liberal Arts Department, 
School of the Art Institute of Chicago, 
112 S Michigan Ave, Chicago, IL 60603}
\altaffiltext{\KASI}{Korea Astronomy and Space Science Institute, Daejeon 305-348, Republic of Korea}
\altaffiltext{\LLNL}{Institute of Geophysics and Planetary Physics, Lawrence
Livermore National Laboratory, Livermore, CA 94551}
\altaffiltext{\Dunlap}{Dunlap Institute for Astronomy \& Astrophysics, University of Toronto, 50 St George St, Toronto, ON, M5S 3H4, Canada}
\altaffiltext{\Toronto}{Department of Astronomy \& Astrophysics, University of Toronto, 50 St George St, Toronto, ON, M5S 3H4, Canada}
\altaffiltext{\BCCP}{Berkeley Center for Cosmological Physics,
Department of Physics, University of California, and Lawrence Berkeley
National Labs, Berkeley, CA 94720}
\altaffiltext{\CTIO}{Cerro Tololo Inter-American Observatory, Casilla 603, La Serena, Chile}

\def\ANL{1}
\def\KICPChicago{2}
\def\PhysicsUChicago{3}
\def\CfA{4}
\def\McGill{5}
\def\UChicago{6}
\def\KIPAC{7}
\def\Stanford{8}
\def\SLAC{9}
\def\AIfA{10}
\def\MIT{11}
\def\Harvard{12}
\def\FNAL{13}
\def\AAUChicago{14}
\def\Munich{15}
\def\ExcellenceCluster{16}
\def\Miss{17}
\def\EFIChicago{18}
\def\NIST{19}
\def\PUC{20}
\def\Caltech{21}
\def\CIFAR{22}
\def\illast{23}
\def\illphy{24}
\def\Berkeley{25}
\def\MPE{26}
\def\UFlorida{27}
\def\Colorado{28}
\def\LeidenObservatory{29}
\def\Davis{30}
\def\LBNL{31}
\def\Arizona{32}
\def\Michigan{33}
\def\Minnesota{34}
\def\Melbourne{35}
\def\STScI{36}
\def\CaseWestern{37}
\def\SAIC{38}
\def\KASI{39}
\def\LLNL{40}
\def\Dunlap{41}
\def\Toronto{42}
\def\BCCP{43}
\def\CTIO{44}

\author{
L.~E.~Bleem\altaffilmark{\ANL,\KICPChicago,\PhysicsUChicago},
B.~Stalder\altaffilmark{\CfA},
T.~de~Haan\altaffilmark{\McGill},
K.~A.~Aird\altaffilmark{\UChicago},
S.~W.~Allen\altaffilmark{\KIPAC,\Stanford,\SLAC},
D.~E.~Applegate\altaffilmark{\AIfA},
M.~L.~N.~Ashby\altaffilmark{\CfA},
M.~Bautz\altaffilmark{\MIT},
M.~Bayliss\altaffilmark{\CfA,\Harvard},
B.~A.~Benson\altaffilmark{\KICPChicago,\FNAL,\AAUChicago},
S.~Bocquet\altaffilmark{\Munich,\ExcellenceCluster},
M.~Brodwin\altaffilmark{\Miss},
J.~E.~Carlstrom\altaffilmark{\ANL,\KICPChicago,\PhysicsUChicago,\AAUChicago,\EFIChicago}, 
C.~L.~Chang\altaffilmark{\ANL,\KICPChicago,\EFIChicago}, 
I.~Chiu\altaffilmark{\Munich,\ExcellenceCluster},
H.~M.~Cho\altaffilmark{\NIST}, 
A.~Clocchiatti\altaffilmark{\PUC},
T.~M.~Crawford\altaffilmark{\KICPChicago,\AAUChicago},
A.~T.~Crites\altaffilmark{\KICPChicago,\AAUChicago,\Caltech},
S.~Desai\altaffilmark{\Munich,\ExcellenceCluster},
J.~P.~Dietrich\altaffilmark{\Munich,\ExcellenceCluster},
M.~A.~Dobbs\altaffilmark{\McGill,\CIFAR},
R.~J.~Foley\altaffilmark{\CfA,\illast,\illphy},
W.~R.~Forman\altaffilmark{\CfA},
E.~M.~George\altaffilmark{\Berkeley,\MPE},
M.~D.~Gladders\altaffilmark{\KICPChicago,\AAUChicago},
A.~H.~Gonzalez\altaffilmark{\UFlorida},
N.~W.~Halverson\altaffilmark{\Colorado},
C.~Hennig\altaffilmark{\Munich,\ExcellenceCluster},
H.~Hoekstra\altaffilmark{\LeidenObservatory}, 
G.~P.~Holder\altaffilmark{\McGill},
W.~L.~Holzapfel\altaffilmark{\Berkeley},
J.~D.~Hrubes\altaffilmark{\UChicago},
C.~Jones\altaffilmark{\CfA},
R.~Keisler\altaffilmark{\KICPChicago,\PhysicsUChicago,\KIPAC,\Stanford},
L.~Knox\altaffilmark{\Davis},
A.~T.~Lee\altaffilmark{\Berkeley,\LBNL},
E.~M.~Leitch\altaffilmark{\KICPChicago,\AAUChicago},
J.~Liu\altaffilmark{\Munich,\ExcellenceCluster},
M.~Lueker\altaffilmark{\Caltech,\Berkeley},
D.~Luong-Van\altaffilmark{\UChicago},
A.~Mantz\altaffilmark{\KICPChicago},
D.~P.~Marrone\altaffilmark{\Arizona},
M.~McDonald\altaffilmark{\MIT},
J.~J.~McMahon\altaffilmark{\Michigan},
S.~S.~Meyer\altaffilmark{\KICPChicago,\PhysicsUChicago,\AAUChicago,\EFIChicago},
L.~Mocanu\altaffilmark{\KICPChicago,\AAUChicago},
J.~J.~Mohr\altaffilmark{\Munich,\ExcellenceCluster,\MPE},
S.~S.~Murray\altaffilmark{\CfA},
S.~Padin\altaffilmark{\KICPChicago,\AAUChicago,\Caltech},
C.~Pryke\altaffilmark{\Minnesota}, 
C.~L.~Reichardt\altaffilmark{\Berkeley,\Melbourne},
A.~Rest\altaffilmark{\STScI},
J.~Ruel\altaffilmark{\Harvard},
J.~E.~Ruhl\altaffilmark{\CaseWestern}, 
B.~R.~Saliwanchik\altaffilmark{\CaseWestern}, 
A.~Saro\altaffilmark{\Munich},
J.~T.~Sayre\altaffilmark{\CaseWestern}, 
K.~K.~Schaffer\altaffilmark{\KICPChicago,\EFIChicago,\SAIC}, 
T.~Schrabback\altaffilmark{\AIfA},
E.~Shirokoff\altaffilmark{\Caltech,\Berkeley}, 
J.~Song\altaffilmark{\Michigan,\KASI},
H.~G.~Spieler\altaffilmark{\LBNL},
S.~A.~Stanford\altaffilmark{\Davis,\LLNL},
Z.~Staniszewski\altaffilmark{\Caltech,\CaseWestern},
A.~A.~Stark\altaffilmark{\CfA}, 
K.~T.~Story\altaffilmark{\KICPChicago,\PhysicsUChicago},
C.~W.~Stubbs\altaffilmark{\CfA,\Harvard}, 
K.~Vanderlinde\altaffilmark{\Dunlap,\Toronto},
J.~D.~Vieira\altaffilmark{\illast,\illphy},
A. Vikhlinin\altaffilmark{\CfA},
R.~Williamson\altaffilmark{\KICPChicago,\AAUChicago,\Caltech}, 
O.~Zahn\altaffilmark{\Berkeley,\BCCP},
and
A.~Zenteno\altaffilmark{\Munich,\CTIO}
}

%% file: field_tablevfeb28.tex
\textsc{ra5h30dec$-$55} & 82.7 & $-$55.0 & 82.9 & 38.2 & 12.8 & 37.0 & 2008,2011 \\
\textsc{ra23h30dec$-$55} & 352.5 & $-$55.0 & 100.2 & 36.9 & 11.7 & 35.0 & 2008,2010 \\
\textsc{ra21hdec$-$60} & 315.0 & $-$60.0 & 147.6 & 35.5 & 15.0 & 58.1 & 2009 \\
\textsc{ra3h30dec$-$60} & 52.5 & $-$60.0 & 222.6 & 35.7 & 15.7 & 59.0 & 2009 \\
\textsc{ra21hdec$-$50} & 315.0 & $-$50.0 & 190.0 & 40.7 & 17.7 & 65.7 & 2009 \\
\textsc{ra4h10dec$-$50} & 62.5 & $-$50.0 & 155.5 & 30.9 & 14.4 & 59.5 & 2010 \\
\textsc{ra0h50dec$-$50} & 12.5 & $-$50.0 & 156.2 & 36.8 & 16.1 & 64.2 & 2010 \\
\textsc{ra2h30dec$-$50} & 37.5 & $-$50.0 & 155.7 & 35.1 & 15.2 & 58.5 & 2010 \\
\textsc{ra1hdec$-$60} & 15.0 & $-$60.0 & 145.9 & 34.6 & 15.6 & 60.1 & 2010 \\
\textsc{ra5h30dec$-$45} & 82.5 & $-$45.0 & 102.7 & 39.1 & 17.7 & 72.7 & 2010 \\
\textsc{ra6h30dec$-$55} & 97.5 & $-$55.0 & 83.3 & 35.6 & 15.7 & 65.2 & 2011 \\
\textsc{ra3h30dec$-$42.5} & 52.5 & $-$42.5 & 166.8 & 34.0 & 15.4 & 62.4 & 2011 \\
\textsc{ra23hdec$-$62.5} & 345.0 & $-$62.5 & 70.5 & 35.9 & 15.8 & 62.2 & 2011 \\
\textsc{ra21hdec$-$42.5} & 315.0 & $-$42.5 & 111.2 & 36.7 & 16.5 & 67.0 & 2011 \\
\textsc{ra1hdec$-$42.5} & 15.0 & $-$42.5 & 108.6 & 35.4 & 15.3 & 62.9 & 2011 \\
\textsc{ra22h30dec$-$55} & 337.5 & $-$55.0 & 83.6 & 37.3 & 16.3 & 67.2 & 2011 \\
\textsc{ra23hdec$-$45} & 345.0 & $-$45.0 & 204.5 & 35.0 & 15.6 & 64.4 & 2011 \\
\textsc{ra6h30dec$-$45} & 97.5 & $-$45.0 & 102.8 & 35.8 & 15.6 & 67.0 & 2011 \\
\textsc{ra6hdec$-$62.5} & 90.0 & $-$62.5 & 68.7 & 34.9 & 15.9 & 67.9 & 2011 \\

%% file: tab_missed_2500d.tex
SPT-CL J0003$-$5253 &       0.8237 &     $-$52.8970 &     5.37 &     2.75 & N & N & - \\ 
SPT-CL J0115$-$5959 &      18.8096 &     $-$59.9887 &     5.91 &     0.25 & Y & Y & - \\ 
SPT-CL J0205$-$4125 &      31.3274 &     $-$41.4224 &     7.05 &     1.50 & N & N & - \\ 
SPT-CL J0222$-$5335 &      35.6926 &     $-$53.5970 &     5.16 &     0.25 & N & N & - \\ 
SPT-CL J0245$-$5302 &      41.3805 &     $-$53.0359 &    15.95 &     0.50 & Y & Y & ACO S0295\tablenotemark{a} \\ 
SPT-CL J0321$-$4515 &      50.3200 &     $-$45.2556 &     5.88 &     1.50 & N & N & - \\ 
SPT-CL J0334$-$6008 &      53.6899 &     $-$60.1497 &     7.11 &     1.25 & N & Y & - \\ 
SPT-CL J0336$-$4037 &      54.0689 &     $-$40.6314 &     9.94 &     0.75 & Y & Y & ACO 3140\tablenotemark{a} \\
SPT-CL J0434$-$5727 &      68.6315 &     $-$57.4523 &     5.06 &     0.75 & Y & Y & - \\ 
SPT-CL J0440$-$4744 &      70.2431 &     $-$47.7370 &     5.18 &     1.25 & N & Y & - \\ 
SPT-CL J0442$-$5905 &      70.6496 &     $-$59.0929 &     6.27 &     0.25 & N & N & - \\ 
SPT-CL J0616$-$5227 &      94.1430 &     $-$52.4546 &     7.81 &     0.75 & Y & Y & 	ACT-CL J0616$-$5227\tablenotemark{b} \\ 
SPT-CL J2104$-$4120 &     316.0754 &     $-$41.3475 &     8.97 &     2.25 & Y & Y & ACO 3739\tablenotemark{a} \\
SPT-CL J2142$-$6419 &     325.7063 &     $-$64.3224 &     9.44 &     0.25 & N & N & - \\ 
SPT-CL J2154$-$5952 &     328.6973 &     $-$59.8821 &     7.10 &     0.50 & N & N & - \\ 
SPT-CL J2154$-$5936 &     328.7003 &     $-$59.6068 &     6.31 &     0.50 & Y & Y & - \\ 
SPT-CL J2246$-$5244 &     341.5844 &     $-$52.7430 &     5.67 &     3.00 & Y & Y & ACO 3911\tablenotemark{a} \\
SPT-CL J2300$-$5100 &     345.1034 &     $-$51.0126 &     5.22 &     2.50 & N & N & - \\ 
SPT-CL J2347$-$6246 &     356.8065 &     $-$62.7693 &     6.63 &     0.25 & N & N & ACO 4036\tablenotemark{a} \\ 

%% file: final_cluster_table.tex
SPT-CL J0000$-$4356 & $0.0663 $ & $ -43.9494 $ & $ 5.92 $ & $ 0.25 $ & $ 76\pm 21$ & $1.00\pm 0.11$ & $3.27\pm 0.71 $ & $ 2$ & \\
SPT-CL J0000$-$5748$^{*}$ & $0.2499 $ & $ -57.8064 $ & $ 8.49 $ & $ 0.50 $ & $ 82\pm 12$ & $0.702^{+}$ & $4.56\pm 0.80 $ & $ 2$ & a\\
SPT-CL J0001$-$4842 & $0.2768 $ & $ -48.7132 $ & $ 5.69 $ & $ 1.25 $ & $ 77\pm 12$ & $0.30\pm 0.04$ & $3.96\pm 0.87 $ & $ 5$ & \\
SPT-CL J0001$-$4024 & $0.3610 $ & $ -40.4108 $ & $ 5.42 $ & $ 0.75 $ & $ 59\pm 12$ & $0.85\pm 0.07$ & $3.12\pm 0.75 $ & $ 2,10$ & \\
SPT-CL J0001$-$6258 & $0.4029 $ & $ -62.9808 $ & $ 4.69 $ & $ 1.50 $ & $ 51\pm 16$ & $0.21\pm 0.02$ & $3.27\pm 0.91 $ & $ 3$ & b,c\\
SPT-CL J0001$-$5440 & $0.4059 $ & $ -54.6697 $ & $ 5.69 $ & $ 1.00 $ & $ 60\pm 12$ & $0.73\pm 0.06$ & $3.54\pm 0.79 $ & $ 1,10$ & \\
SPT-CL J0002$-$5557 & $0.5138 $ & $ -55.9621 $ & $ 5.20 $ & $ 0.25 $ & $ 54\pm 18$ & $1.14\pm 0.08$ & $2.47\pm 0.60 $ & $ 3,10$ & \\
SPT-CL J0002$-$5224 & $0.6433 $ & $ -52.4092 $ & $ 4.67 $ & $ 1.00 $ & $ 48\pm 12$ & $>0.71$ & $$\nodata$ $ & $ 2$ & \\

%% file: tab_other.tex
SPT-CL J0010$-$5112 & ACO S0013, 1 & 1 & 0.17$\pm$ 0.02 & $-$ \\
SPT-CL J0013$-$4621 & [BM78] 210, 2 & 1,2 & 0.18$\pm$ 0.05 & $-$ \\
SPT-CL J0027$-$5015 & Str 0025$-$505, 3 & 1,3,4,5,6 & 0.145 & 0.1448,[4] \\
SPT-CL J0036$-$4411 & SpARCS J003645$-$441050, 7 & 7 & 0.869 & 0.869,[7] \\
SPT-CL J0041$-$4428 & ACO S0067, 1 & 1 & 0.33$\pm$ 0.02 & $-$ \\
SPT-CL J0051$-$4834 & Str 0049$-$489, 3 & 1,3 & 0.187 & 0.1873,[8] \\
SPT-CL J0102$-$4915 & ACT-CL J0102$-$4915, 9 & 6,9,10 & 0.870 & 0.8701,[11] \\
SPT-CL J0108$-$4341 & ACO 2873, 1 & 1 & 0.19$\pm$ 0.02 & $-$ \\